\documentclass[%
 reprint,
superscriptaddress,
 amsmath,amssymb,
 aps,
prl,
floatfix,
]{revtex4-1}

\usepackage{graphicx}% Include figure files
\usepackage{dcolumn}% Align table columns on decimal point
\usepackage{bm}% bold math
\usepackage[colorlinks=true,allcolors=blue]{hyperref}% add hypertext capabilities
\usepackage{array,booktabs,tabularx}
\usepackage[caption=false]{subfig}
\usepackage[USenglish]{babel}
\usepackage{braket}
\usepackage{xcolor}
\usepackage[normalem]{ulem}
\usepackage{ulem}

\newcommand{\beq}{\begin{equation}}
\newcommand{\eeq}{\end{equation}}
\newcommand{\beqn}{\begin{eqnarray}}
\newcommand{\eeqn}{\end{eqnarray}}
\newcommand{\scharho}{\rho_{\boldsymbol{\mathcal{R}}\boldsymbol{\Phi}}}
\newcommand{\rscha}{\boldsymbol{\mathcal{R}}}
\newcommand{\phischa}{\boldsymbol{\Phi}}

\usepackage{etoolbox} % Provides ability to hook into thebibliography
\newcounter{firstbib} % New counter: keeps track of the next first number

\AtBeginEnvironment{thebibliography}{
}

\apptocmd{\thebibliography}{
  \setcounter{NAT@ctr}{\value{firstbib}} % Use this for natbib and revtex
}{}{}  % These args are for errors handling: see etoolbox docs

\AtEndEnvironment{thebibliography}{
  \setcounter{firstbib}{\value{NAT@ctr}}
}

\begin{document}
\selectlanguage{USenglish}
\preprint{APS/123-QED}

%\title{Mechanically stable graphene with quadratic out-of-plane acoustic modes}
%\title{Flexural acoustic phonons and the origin of sound propagation in graphene}
\title{Bending rigidity, sound propagation and ripples in flat graphene}

\author{Unai Aseginolaza}
%\email{uaseguinolaz001@ehu.eus}
\affiliation{Centro de F\'isica de Materiales CFM, CSIC-UPV/EHU, Paseo Manuel de
             Lardizabal 5, 20018 Donostia/San Sebasti\'an, Spain}
%\affiliation{Donostia International Physics Center
%             (DIPC), Manuel Lardizabal pasealekua 4, 20018 Donostia, Basque Country, Spain}
\affiliation{Fisika Aplikatua Saila, 
             University of the Basque Country (UPV/EHU), Europa Plaza 1, 20018 Donostia/San Sebasti\'an, Spain}
\affiliation{Basic Sciences Department, Faculty of Engineering, Mondragon Unibertsitatea, 20500 Arrasate, Spain}
             
\author{Josu Diego}
%\email{josu.diego@ehu.eus}
\affiliation{Centro de F\'isica de Materiales CFM, CSIC-UPV/EHU, Paseo Manuel de
             Lardizabal 5, 20018 Donostia/San Sebasti\'an, Spain}
\affiliation{Fisika Aplikatua Saila, 
             University of the Basque Country (UPV/EHU), Europa Plaza 1, 20018 Donostia/San Sebasti\'an, Spain}
             
\author{Tommaso Cea}
\affiliation{Department of Physical and Chemical Sciences, Universit\'a degli Studi dell’Aquila, I-67100 L’Aquila, Italy}
\affiliation{Imdea Nanoscience, Faraday 9, 28015 Madrid, Spain}
\affiliation{Dipartimento di Fisica, Universit\`a di Roma La Sapienza, Piazzale Aldo Moro 5, I-00185 Roma, Italy} 
\affiliation{Graphene Labs, Fondazione Instituto Italiano di Tecnologia, Italy}

\author{Raffaello Bianco}
\affiliation{Centro de F\'isica de Materiales CFM, CSIC-UPV/EHU, Paseo Manuel de
             Lardizabal 5, 20018 Donostia/San Sebasti\'an, Spain}
\affiliation{Ru\dj er Bo\v{s}kovi\'c Institute, 10000 Zagreb, Croatia}
\affiliation{Dipartimento di Scienze Fisiche, Informatiche e Matematiche, Universit\`a di Modena e Reggio Emilia, Via Campi 213/a I-41125 Modena, Italy}
\affiliation{Centro S3, Istituto Nanoscienze-CNR, Via Campi 213/a, I-41125 Modena, Italy}
             
\author{Lorenzo Monacelli}
\affiliation{Theory and Simulation of Materials (THEOS), École Polytechnique Fédérale de Lausanne, CH-1015 Lausanne, Switzerland}

\author{Francesco Libbi}
\affiliation{Theory and Simulation of Materials (THEOS), École Polytechnique Fédérale de Lausanne, CH-1015 Lausanne, Switzerland}

\author{Matteo Calandra}
\affiliation{Dipartimento di Fisica, Università di Trento, Via Sommarive 14, 38123 Povo, Italy.}
\affiliation{Sorbonne Universit\'es, CNRS, Institut des Nanosciences de Paris, UMR7588, F-75252, Paris, France}
\affiliation{Graphene Labs, Fondazione Instituto Italiano di Tecnologia, Italy}

\author{Aitor Bergara}
%\email{a.bergara@ehu.eus}
\affiliation{Centro de F\'isica de Materiales CFM, CSIC-UPV/EHU, Paseo Manuel de
             Lardizabal 5, 20018 Donostia/San Sebasti\'an, Spain}
\affiliation{Donostia International Physics Center
             (DIPC), Manuel Lardizabal pasealekua 4, 20018 Donostia/San Sebasti\'an, Spain}
\affiliation{Departamento de F\'isica and EHU Quantum Center,  University of the Basque Country (UPV/EHU), 48080 Bilbao, 
             Basque Country, Spain}
             
\author{Francesco Mauri}
\affiliation{Dipartimento di Fisica, Universit\`a di Roma La Sapienza, Piazzale Aldo Moro 5, I-00185 Roma, Italy} 
\affiliation{Graphene Labs, Fondazione Instituto Italiano di Tecnologia, Italy}

\author{Ion Errea}
%\email{ion.errea@ehu.eus}
\affiliation{Centro de F\'isica de Materiales CFM, CSIC-UPV/EHU, Paseo Manuel de
             Lardizabal 5, 20018 Donostia/San Sebasti\'an, Spain}
\affiliation{Fisika Aplikatua Saila, 
             University of the Basque Country (UPV/EHU), Europa Plaza 1, 20018 Donostia/San Sebasti\'an, Spain}             
\affiliation{Donostia International Physics Center
             (DIPC), Manuel Lardizabal pasealekua 4, 20018 Donostia/San Sebasti\'an, Spain}

\date{\today}% It is always \today, today,

                             % Classification Scheme./
\maketitle

%*********************************
%The discovery of graphene~\cite{novoselov2004electric,novoselov2005two,novoselov2005twoo,zhang2005experimental} proved the existence of 2D materials and launched their science and technology. Graphene is already a reality in different industrial products~\cite{Kong2019} that benefit from its fantastic properties. In particular, its mechanical and thermal properties are crucial for many of its current and future applications. For instance, graphene's uneven strength, stiffness, and lightness~\cite{lee2008measurement} have been used to make stronger but lighter macroscopic objects, such as tennis rackets, shoes, and so on. Graphene, due to its very high thermal conductivity~\cite{ghosh2008extremely}, has been already incorporated into electronic devices like mobile phones for efficient heat dissipation. 
%*********************************

\textbf{
Despite many of the applications of graphene rely on its uneven stiffness and high thermal conductivity, the mechanical properties of graphene, and in general of all 2D materials, are still elusive. The harmonic theory predicts a quadratic dispersion for the flexural acoustic vibrational mode, which leads the unphysical result that long wavelength in-plane acoustic modes decay before vibrating one period, preventing the propagation of sound. The robustness of the quadratic dispersion has been questioned by arguing that the anharmonic phonon-phonon interaction linearizes it. However, this implies a divergent bending rigidity in the long wavelength limit not reproduced experimentally. Here we show that rotational symmetry protects the quadratic flexural dispersion against phonon-phonon interactions and that, consequently, the bending stiffness is non-divergent irrespective of the temperature. Our non-perturbative anharmonic calculations also determine that sound propagation coexists with a quadratic dispersion. We also show that the temperature dependence of the height fluctuations of the membrane, known as ripples, is fully determined by thermal or quantum fluctuations, but without the anharmonic suppression of their amplitude previously assumed. The universality of our conclusions reconcile experimental evidence and theory not just in graphene, but all 2D materials.
}

The theoretical comprehension of the mechanical properties of 2D materials and membranes, which affect their acoustic and thermal properties, is one of the oldest problems in  condensed matter physics, dating back to the times in which the possibility of having 2D crystalline order was questioned~\cite{landau_statistical_physics,mermin1968crystalline}. Even if the discovery of graphene and other 2D materials~\cite{novoselov2004electric,novoselov2005twoo,Meyer2007} put aside this question, the understanding of how these materials can propagate sound, what is their bending rigidity, and the amplitude of their ripples are still under strong debate~\cite{refId0,Fasolino2007,gazit_PRB2009,gazit_PRE2009,sanjose_PRL2011,Bonilla_JSM2012,guinea_ledoussal_PRB2014,gonzalez_PRB2014,ruiz_JSM2015,gornyi_PRB2015,garcia_PRB2016,bonilla_PRB2016,gornyi_JETP2016,garcia_PRE2017,cea_cm2019,cea_cm2019_2,silva_PRB2020}. No unifying picture has emerged yet.

%Many of the applications of graphene are based on its uneven stiffness and high thermal conductivity~\cite{Kong2019,lee2008measurement,ghosh2008extremely}. However, the mechanical, thermal, and acoustic properties of graphene, and in general of any 2D material, are far from been understood yet. Even the possibility of having crystalline order in 2D has been long questioned due to the diverging atomic displacements as a function of the sample size calculated in classical references~\cite{landau_statistical_physics,mermin1968crystalline}. Experimentally, crystalline order has been observed in suspended graphene~\cite{Meyer2007}, although it shows ripples that seem to be intrinsic according to many theoretical studies~\cite{refId0,Fasolino2007,gazit_PRB2009,gazit_PRE2009,sanjose_PRL2011,Bonilla_JSM2012,guinea_ledoussal_PRB2014,gonzalez_PRB2014,ruiz_JSM2015,gornyi_PRB2015,garcia_PRB2016,bonilla_PRB2016,gornyi_JETP2016,garcia_PRE2017,cea_cm2019,cea_cm2019_2,silva_PRB2020}.

Most of the theoretical problems are caused by the quadratic dispersion of the acoustic flexural out-of-plane (ZA) mode that is obtained in the  harmonic approximation. Such a quadratic dispersion also implies the unphysical result that graphene and other 2D membranes do not propagate sound. Indeed, the phonon linewidths of the in-plane acoustic longitudinal (LA) and transverse (TA) phonons calculated perturbatively from the harmonic result do not vanish in the long wavelength limit~\cite{paulatto2013anharmonic}, precisely, because of the quadratic dispersion of the ZA modes~\cite{bonini2012acoustic}. This yields the conclusion that phonons having sufficiently small momentum do not live long enough for vibrating one period and, thus, the quasiparticle picture is lost together with the propagation of sound. 

It has been argued~\cite{wang2016anharmonic,los2009scaling,katsnelson2013graphene,zakharchenko2009finite,mariani2008flexural,amorim2014thermodynamics,de2012bending,PhysRevB.91.134302,PhysRevB.96.094302} that the anharmonic coupling between in-plane and out-of-plane phonon modes renormalizes the dispersion of the ZA phonon modes, providing it with a linear term at small momenta that somewhat cures the pathologies. It has long been assumed~\cite{refId0,PhysRevLett.69.1209} as well that the out-of-plane vibrational frequency of any continuous membrane acquires a linear term at small wavevectors once anharmonic interactions are included.
The linear term stiffens the membrane and consequently suppresses the amplitude of its ripples, which is usually studied from the height correlation function in momentum space, $\langle |h(\mathbf{q})|^2\rangle$. In the harmonic approximation it scales as $\langle |h(\mathbf{q})|^2\rangle\sim q^{-4}$ and it is corrected to $q^{-4+\eta}$, with $\eta\sim 0.80-0.85$, when the ZA modes is linearized~\cite{refId0,PhysRevLett.69.1209,mariani2008flexural,amorim2014thermodynamics,de2012bending}. 
Since the bending rigidity scales as $\langle |h(\mathbf{q})|^2\rangle q^4$~\cite{refId0}, this interpretation implies that the bending stiffness of all membranes and 2D materials diverges  in the long wavelength limit, yielding the dubious interpretation that the larger the membrane, the stiffer it becomes. The experimental confirmation of these ideas is challenging due to the difficulties in measuring the bending rigidity of graphene~\cite{Blees2015,Lindahl2012} and the substrate effects on the dispersion of the ZA modes measured with hellium diffraction~\cite{al2016acoustic,al2015helium,al2018resolving}. However, the fact that independent experiments~\cite{al2015helium,tomterud2022temperature} find consistent values of the bending rigidity questions this picture. 

%In this letter, by performing non-perturbative anharmonic calculations on graphene using both atomistic calculations and a membrane model, we convincingly show that a quadratic dispersion of the ZA mode in unstrained graphene, and any other 2D membrane, is actually expected and that it is compatible with well-defined sound waves. We also show that the bending stiffness of graphene is barely unaffected by phonon-phonon interactions. Our results are in stark contrast to the previously assumed behavior of membranes because we fully preserve rotational invariance and do not work with a low-energy model that breaks this symmetry.   

The quadratic dispersion expected for the ZA mode in the harmonic approximation is imposed by symmetry. In this case phonon frequencies are obtained diagonalizing the $\phi_{ab}/\sqrt{M_aM_b}$ dynamical matrix, where $a$ and $b$ represent both atom and Cartesian indices, $M_a$ is the mass of atom $a$, and $\phi_{ab}= \left[\frac{\partial V}{\partial R_a \partial R_b}\right]_{\scriptstyle{0}}$ are the second-order force constants obtained as the second-order derivatives of the Born-Oppenheimer potential $V$ with respect to atomic positions $\boldsymbol{R}$ calculated at the positions that minimize $V$. Rotational symmetry, together with the fact that in a strictly two-dimensional system force constants involving an in-plane and an out-of-plane displacement vanish, makes the ZA mode acquire a quadratic dispersion close to zone center~\cite{katsnelson2013graphene}. Phonons expected experimentally, however, should be calculated from the imaginary part of the phonon Green's function that includes anharmonic effects~\cite{bianco2017second}. For low energy modes, such as the ZA mode, dynamical effects can be safely neglected. In this limit the phonon peaks coincide with the eigenvalues of the free energy Hessian $[\frac{\partial F}{\partial \mathcal{R}_a \partial \mathcal{R}_b}]_{0}/\sqrt{M_aM_b}$, where $F$ is the anharmonic free energy, $\rscha$ the average ionic positions, and the derivative is taken at the positions that minimize $F$~\cite{bianco2017second}. This raises a formidable remark that has remained unnoticed thus far: as $F$ and $V$ obey the same symmetry properties, a quadratic dispersion should be expected for the ZA mode not only in the harmonic limit, also when anharmonic interactions are considered.   

We dig into this point by accounting for anharmonicity beyond perturbation theory within the self-consistent harmonic approximation (SCHA). The SCHA is applied both in its stochastic implementation~\cite{bianco2017second,errea2014anharmonic,monacelli2018pressure} by making use of a machine learning atomistic potential~\cite{rowe2018development} and with  a membrane continuum Hamiltonian. The SCHA is a variational method that minimizes the free energy of the system 
\beq
F=\langle T + V + \frac{1}{\beta}\ln \scharho  \rangle_{\scharho}  
\label{sscha_f}
\eeq
with respect to a density matrix $\scharho$ parametrized with centroid positions $\rscha$ and auxiliary force constants $\phischa$ (bold symbols represent vectors or tensors in compact notation). In Eq. \eqref{sscha_f} $T$ is the ionic kinetic energy, $\beta$ the inverse temperature, and $\langle O \rangle_{\scharho} = \textrm{tr}[\scharho O]$ ($O$ is any operator). We call \emph{auxiliary} the phonon frequencies obtained diagonalizing the $\Phi_{ab}/\sqrt{M_aM_b}$ matrix. These frequencies include non-perturbative anharmonic corrections as they result from the variational minimization of $F$ that fully includes $V$. However, phonons probed experimentally are related to the peaks in the imaginary part of the analytical continuation of the interacting Green's function $G_{ab}(\omega+i\delta)$~\cite{bianco2017second,monacelli2021time,lihm2021gaussian}, which can be calculated from the  
\beq
G^{-1}_{ab}(i\Omega_n) = G^{-1(S)}_{ab}(i\Omega_n) -\Pi_{ab}(i\Omega_n) 
\label{green-dyson}
\eeq
Dyson's equation, where $\Omega_n$ are bosonic Matsubara's frequencies. In Eq. \eqref{green-dyson}, $G^{-1(S)}_{ab}(i\Omega_n)=(i\Omega_n)^2\delta_{ab}-\Phi_{ab}/\sqrt{M_aM_b}$ is the non-interacting Green's function formed by the auxiliary phonons and $\boldsymbol{\Pi}(i\Omega_n)$ is the phonon-phonon interaction self-energy, which we estimate within the SCHA (see Methods). The peaks in the imaginary part of $G_{ab}(\omega+i\delta)$ determine the frequencies and linewidths of the \emph{physical} phonons. In the static $\omega=0$ limit the peaks coincide with the eigenvalues of the free energy Hessian.

\begin{figure*}[ht]
\includegraphics[width=1\linewidth]{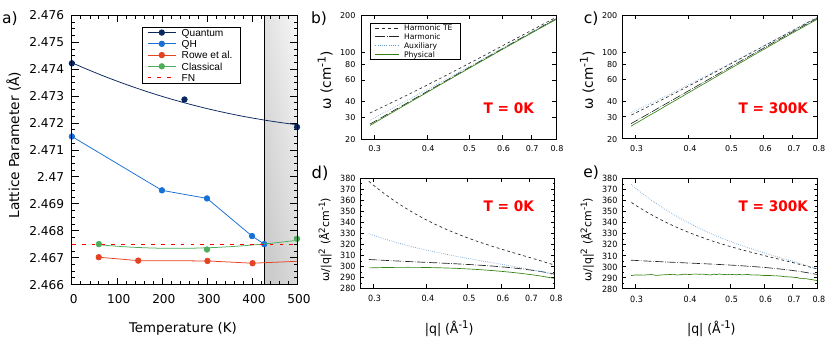}
\caption{(a) Lattice parameter of graphene as a function of temperature obtained with the SCHA using a machine learning atomistic potential. Both quantum and classical  calculations are included. The temperature-independent frozen nuclei (FN) result corresponds to the lattice parameter that minimizes the Born Oppenheimer potential $V$. The MD results obtained by Rowe et al.~\cite{rowe2018development} are included. The lattice parameter calculated in the quasiharmonic (QH) approximation is also included. In the grey zone harmonic phonons become unstable breaking down the quasiharmonic approximation. (b)-(e) Harmonic ZA phonon spectra together with the SCHA auxiliary phonons and the physical phonons obtained from the peaks of the Green's function in Eq. \eqref{green-dyson} at 0 K (b) and 300 K (c). Panels (d) and (e) show the bending rigidity, defined as the frequency 
divided by the squared momentum. In the panels the dispersion corresponds to the $\Gamma$M direction. For reference, the M point is at $1.4662$ $\AA^{-1}$ at $0$ K and at $1.4671$ $\AA^{-1}$ at $300$ K. The harmonic result (solid black) is computed at the lattice parameter that minimizes $V$, while the other results include thermal expansion. The dashed black lines correspond to harmonic calculations including thermal expansion (TE).}
\label{spectrum}
\end{figure*}

In order to preserve rotational symmetry, we make sure that the lattice parameter in our calculations sets the SCHA stress tensor~\cite{monacelli2018pressure} to zero at each temperature. The lattice parameter calculated in this way includes anharmonic effects as well as the effect of quantum and thermal fluctuations. All the phonon spectra shown in this work  obtained with the atomistic potential are calculated with the lattice parameter that gives a null stress at each temperature. The harmonic spectra on the contrary are always calculated at the lattice parameter that minimizes $V$. The temperature dependence of the lattice parameter is shown in Fig. \ref{spectrum}. We include the molecular dynamics (MD) results of Rowe et al. obtained with the same potential~\cite{rowe2018development}, which do not account for quantum effects. For comparison, we also include SCHA calculations in the classical limit, by making $\hbar=0$ in $\scharho$, and within the quasiharmonic (QH) approximation. 
%We expand our calculations up to 500 K, as below this temperature corrugations seem negligible~\cite{pozzo2011thermal,zakharchenko2009finite}. The centroid positions in the SCHA calculations are always in the plane. 
Our quantum calculations correctly capture the negative thermal expansion of graphene that has been estimated in previous theoretical works~\cite{rowe2018development,zakharchenko2009finite}. Our SCHA result shows a larger lattice parameter than the classical result. This is not surprising as classical calculations neglect quantum fluctuations and, consequently, underestimate the fluctuations associated to the high-energy optical modes (the highest energy phonon modes require temperatures of around 2000 K to be thermally populated). This remarks the importance of considering quantum effects in the evaluation of thermodynamic properties of graphene. Our classical results and the MD calculations of Rowe et al.~\cite{rowe2018development} are in agreement at low temperatures.
%d, but disagree at high temperatures. While our calculations approach the quantum result at high temperatures as expected, MD calculations estimate a temperature-independent lattice parameter. We attribute this difference to the difficulties of MD to reach ergodic conditions with respect to the height fluctuations imposed by the low-energy ZA flexural mode, which yields a corrugated average position of the membrane. On the contrary, in our SCHA simulations the centroids are always in the plane. 

%(within their error of $0.0005$ $\AA$). It is worth noting that the quasiharmonic approximation is not valid to calculate the thermal expansion of graphene due to the harmonic imaginary phonon frequencies that appear close to $\Gamma$ for the ZA mode already at $500$ K. In the shadowed region the QH is in principle not valid, even if it has been used with coarse integration grids that avoid the instabilities~\cite{PhysRevB.71.205214}.

In Figs. \ref{spectrum} (b)-(e) we compare the harmonic phonon spectra with the auxiliary phonons as well as with the spectra obtained from the peaks in the imaginary part of the interacting Green's function, the physical phonons. The main conclusion is that while the dispersion of the ZA modes obtained from $\phischa$ is linearized, the physical phonons become close to a quadratic dispersion and approach the harmonic dispersion, as expected by symmetry in the static limit. This is very clear in Fig. \ref{spectrum}(d) and (e), where we show that the bending rigidity, defined as the frequency divided by the squared momentum, is independent of the wavevector at any temperature. This suggests that the bending rigidity is barely affected by interactions, in contradiction to the broadly assumed result that it diverges  at small momentum in membranes due to thermal fluctuations~\cite{refId0}. 
%The auxiliary ZA frequencies suffer a blue-shift with respect to the harmonic ones, but are red-shifted once the physical phonons are calculated. Both shifts are bigger when the temperature is increased, but the quadratic behavior of the ZA modes is always recovered regardless of the temperature. 
%In Fig. \ref{spectrum} (e) we show the temperature dependence of the frequency of the optical $E_{2g}$ mode visible in Raman experiments, which red-shifts with increasing temperature in agreement with experiments~\cite{calizo2007temperature,kagi1994proper,tan1999intrinsic} and previous theoretical works~\cite{bonini2007phonon}. The experimental results in graphene~\cite{linas2015interplay} lay between the shift calculated considering thermal expansion and without considering it, which is the expected behavior due to the presence of the substrate.

\begin{figure}[ht]
\includegraphics[width=1.0\linewidth]{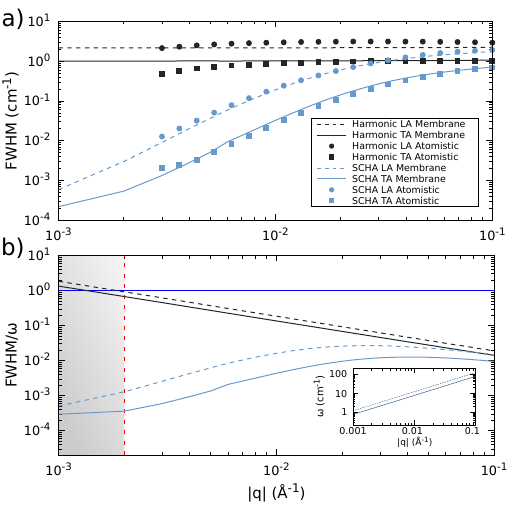}
\caption{(a) Linewidths (full width at half maximum) of LA and TA phonon modes at $300$ K calculated within perturbation theory on top of the harmonic result and within the SCHA following Eq. \eqref{green-dyson}. Squares and circles are calculated with the atomistic potential and lines correspond to calculations within the membrane model. Our harmonic results are in good agreement with other theoretical calculations~\cite{paulatto2013anharmonic,bonini2012acoustic}. (b) FWHM divided by the phonon frequency in the membrane model. In the inset we show the phonon frequencies in the same momentum range. The grey zone corresponds to the region where the fraction FWHM$_{LA}/\omega_{LA}$ is bigger than one in the harmonic case.}
\label{problems}
\end{figure}

Even if the anharmonic correction to the phonon spectra may look small in Fig. \ref{spectrum}, it has a huge impact on the acoustic properties of graphene. As shown in Fig. \ref{problems}, the SCHA non-perturbative calculation based on Eq. \eqref{green-dyson} dramatically changes the linewidth of the LA and TA modes at small momenta by making them smaller as momentum decreases, in clear contrast to the perturbative calculation obtained on top of the harmonic result. This happens thanks to the linearization of the auxiliary flexural phonons that form the non-interacting Green's function and enter in Dyson's equation. When the ratio between the full-width at half-maximum (FWHM) and the frequency of the mode is approximately 1, the quasiparticle picture is lost. This value is reached in the 0.001-0.002 $\AA^{-1}$ momentum range in the harmonic case. However, when the linewidth is calculated within the SCHA, the ratio never gets bigger than 0.05. These results recover the quasiparticle picture for in-plane acoustic modes at any wavevector, guaranteeing that graphene always propagates sound. The momentum range for which the quasiparticle picture is lost in the harmonic approximation can be reached experimentally with Brillouin scattering probes. In fact, for few layer graphene the quasiparticle picture holds in the 0.001-0.002 $\AA^{-1}$ region~\cite{wang2008brillouin}, in agreement with our calculations. We show here that there is no need of strain~\cite{bonini2012acoustic} to have physically well-defined phonon linewidths in graphene.
%It has been argued previously that strain linearizes the ZA dispersion and makes the linewidths of LA and TA modes vanish at small momenta. We show here that strain linearizes the quadratic ZA physical phonon dispersion in graphene, but, more importantly, we also show that there is no need of strain to have physically well-defined phonon linewidths.

\begin{figure}[t]
\includegraphics[width=1.0\linewidth]{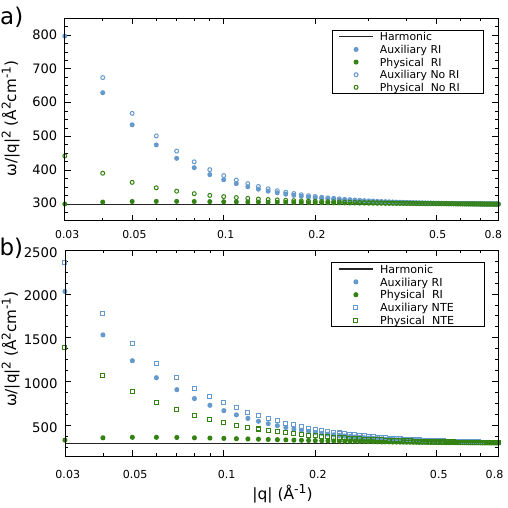}
\caption{(a) Bending rigidity of graphene, defined as the ratio between the frequency of the ZA mode divided by the squared momentum, calculated within the harmonic approximation and within the SCHA auxiliary and physical cases at 0 K in the membrane model. We name rotationally invariant (RI) the results considering the full potential in Eq. \ref{membrane-potential}. We name no rotationally invariant (No RI) the results neglecting the  $\partial_i\boldsymbol{u}\cdot\partial_j\boldsymbol{u}$ term in Eq. \eqref{stress-tensor}. (b) Same results at  $300$ K with the full membrane potential in the rotationally invariant case. We also include the results without considering thermal expansion (NTE).}
\label{membrane}
\end{figure}

In order to obtain results at very small momenta and reinforce the conclusions drawn with the atomistic calculations, we solve the SCHA equations in a continuum membrane Hamiltonian. This model has been widely used in the literature to describe graphene as an elastic membrane as well as to account for the coupling between in-plane and out-of plane acoustic modes~\cite{refId0,PhysRevLett.69.1209,mariani2008flexural,amorim2014thermodynamics,de2012bending}. The most general rotationally invariant continuum potential to describe a free-standing 2D membrane up to the fourth-order in the phonon fields has the following form~\cite{landau1959theory}:
\beqn
 \label{membrane-potential}
 V&=&\frac{1}{2}\int_{\Omega}d^{2}x \left(\kappa(\partial^{2}h)^{2}+C^{ijkl}u_{ij}u_{kl}\right),\\
 \label{stress-tensor}
u_{ij}&=&\frac{1}{2}\left(\partial_i u_j+\partial_j u_i+\partial_i\boldsymbol{u}\cdot\partial_j\boldsymbol{u}+\partial_i h\partial_j h\right).
\eeqn
Here $\boldsymbol{u}(\boldsymbol{x})$ and $h(\boldsymbol{x})$ are the in-plane and out-of-plane displacement fields, respectively,
$u_{ij}$ is the stress tensor,
and $\boldsymbol{x}$ is the 2D position vector in the membrane. $\kappa$ is the harmonic bending rigidity of the membrane, $\Omega$ is the area of the membrane, and the tensor $C^{ijkl}=\lambda\delta^{ij}\delta^{kl}+\mu(\delta^{ik}\delta^{jl}+\delta^{il}\delta^{jk})$ contains the Lam\'e coefficients $\lambda$ and $\mu$ and Kronecker deltas. We have calculated the parameters by fitting them to the atomistic potential, which yields $\lambda=4.3$ eV$\AA^{-2}$, $\mu=9.3$ eV$\AA^{-2}$, $\kappa=1.5$ eV and, $\rho/\hbar^{2}=1097$ eV$^{-1} \AA^{-4}$. This continuum model only accounts for acoustic modes. The harmonic acoustic frequencies given by Eq. \eqref{membrane-potential} are $\omega_{ZA}(q)=\sqrt{\kappa/\rho}q^{2}$, $\omega_{LA}(q)=\sqrt{(\lambda+\mu)/\rho}q$, and $\omega_{TA}(q)=\sqrt{\mu/\rho} q$, $\rho$ being the mass density of the membrane. The thermal expansion is included in this formalism by changing the in-plane derivatives as $\partial_{i}u_{j}\rightarrow \partial_{i}u_{j}+\delta^{ij}\delta a$, with $\delta a=(a-a_{0})/a_{0}$, $a_{0}$ being the lattice parameter that minimizes $V$. 
%We perform the SCHA variational minimization numerically with the potential in Eq. \eqref{membrane-potential}.   

The results obtained in this rotationally invariant membrane are shown in Fig. \ref{membrane}. All conclusions drawn with the atomistic model are confirmed and put in solid grounds. Again the ZA phonons obtained from the auxiliary SCHA force constants get linearized at small momenta. However, when the physical phonons are calculated from the Hessian of the free energy (due to the low frequencies of the ZA modes this static approximation is perfectly valid as shown in the Methods section), the ZA phonon frequencies get on top of the harmonic values recovering a quadratic dispersion. This means that the physical phonons have a quadratic dispersion for small momenta in an unstrained membrane, as it is expected by symmetry, and that the bending rigidity does not increase in the long wavelength limit and is barely affected by interactions. Consequently the bending rigidity that we obtain is around the harmonic value of 1.5 eV, in  good agreement with the consistent experiments by Al Taleb \emph{et al.}~\cite{al2015helium} and T\o{}mterud \emph{et al.}~\cite{tomterud2022temperature}. Fig. \ref{membrane} remarks that accounting correctly for the thermal expansion is crucial to recover the quadratic dispersion of the flexural modes. The validity of the membrane potential is confirmed by calculating the linewidths of the LA and TA modes, which yield consistent results to those obtained with the atomistic potential (see. Fig. \ref{problems}). 

\begin{figure}[t]
\includegraphics[width=1.0\linewidth]{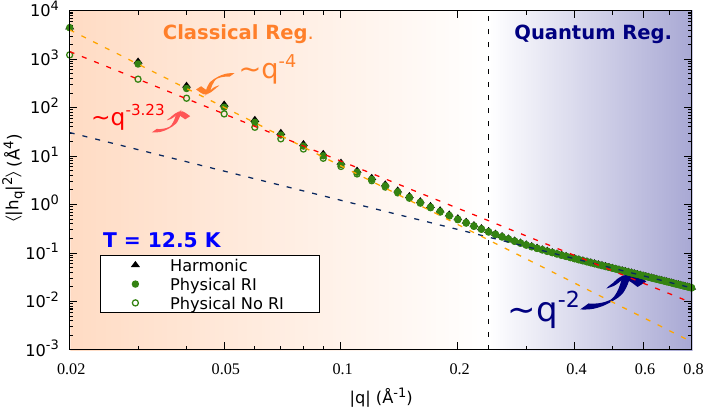}
\caption{Fourier transform of the height-height correlation function at 12.5 K in the membrane model evaluated at different levels of approximation: harmonic (black dots), anharmonic RI result (green filled dots) and anharmonic No RI result (green empty dots). The dashed vertical line specifies the wavevector at which the crossover from classical (orange background) to quantum correlations (violet background) occurs at this temperature. The dashed lines correspond to linear fits with different exponents.}
\label{hh}
\end{figure}

Our results thus upturn the conventional wisdom of 2D membranes~\cite{refId0,PhysRevLett.69.1209,mariani2008flexural,amorim2014thermodynamics,de2012bending}: interactions do not linearize the dispersion of the ZA mode and the bending rigidity does not diverge at small momentum. The main reason for this is that in previous works the $\partial_i\boldsymbol{u}\cdot\partial_j\boldsymbol{u}$ term in the stress tensor, which guarantees rotational invariance, is neglected, unavoidably lowering the power of the ZA phonon frequency to $\sim q^d$ as shown in Fig. \ref{membrane}(a), with $d\sim 1.6$ in our case. 
%Remarkably, this exponent $d$ is in good agreement with the value obtained in the literature in the membrane model 
The amplitude of the height fluctuations or ripples in the long wavelength limit  reflects as well the absence of rotational symmetry in prior calculations. Different calculations within the self-consistent screening approximation or non-perturbative renormalization group theory yield consistent values of $\langle |h(\mathbf{q})|^2\rangle\sim q^{-4+\eta}$, with $\eta\sim 0.80-0.85$~\cite{refId0,PhysRevLett.69.1209,mariani2008flexural,amorim2014thermodynamics,de2012bending,kownacki2009crumpling,roldan2011suppression}. We can estimate $\langle |h(\mathbf{q})|^2\rangle$ within the SCHA in our membrane model by calculating the equal time out-of-plane displacement correlation function, which in the static limit leads to the simple
\begin{equation} 
		\langle |h(\textbf{q})|^2 \rangle = \frac{\left(1+2n_B(\Omega_{ZA}(\textbf{q}))\right)}{2\rho\Omega_{ZA}(\textbf{q})}
  \label{eq:hhq}
\end{equation}
equation (see Methods), where $n_B(\omega)$ is the bosonic occupation factor and $\Omega_{ZA}(\textbf{q})$ the physical flexural phonon frequency coming from the free energy Hessian. The presence of the bosonic occupation completely determines the dependence on $\textbf{q}$ of the correlation function: in the classical limit, when temperature is larger than the frequency of the ZA mode, $\langle |h(\textbf{q})|^2 \rangle \sim \Omega_{ZA}(\textbf{q})^{-2}$, while in the quantum limit, when the ZA mode is unoccupied, $\langle |h(\textbf{q})|^2 \rangle \sim \Omega_{ZA}(\textbf{q})^{-1}$. In the classical regime we recover the $\langle |h(\textbf{q})|^2 \rangle \sim q^{-3.2}$ behavior when we neglect $\partial_i\boldsymbol{u}\cdot\partial_j\boldsymbol{u}$, consistently with previous results (see Fig. \ref{hh}). However, when we keep full rotational invariance, the ZA modes acquires a quadratic dispersion and thus $\langle |h(\textbf{q})|^2 \rangle \sim q^{-4}$, which is the result obtained in the harmonic case. Consequently, anharmonicity does not suppress the amplitude of the ripples in the long wavelength limit, upturning the previous consensus~\cite{refId0,PhysRevLett.69.1209,mariani2008flexural,amorim2014thermodynamics,de2012bending}. It is worth noting that the non rotational invariant membrane deviates from the $q^{-4}$ power law below a critical wave number~\cite{roldan2011suppression}.  

The crossover between the regimes in which thermal and quantum fluctuations determine the ripples (see Fig. \ref{hh}) is in very good agreement with the conclusions drawn with atomistic path-integral Monte Carlo simulations (PIMC) of freestanding graphene~\cite{pimc}. This crossover occurs at different wave numbers depending on the temperature, basically when $\hbar \Omega_{ZA}(\textbf{q}) \sim k_B T$. However, atomistic classical Monte Carlo and MD simulations have estimated $\langle |h(\textbf{q})|^2 \rangle$ for small wave numbers in the order of $q\sim 0.01\AA^{-1}$ finding a scaling law not far from the $q^{-3.2}$ obtained in the membrane model when rotational symmetry is broken~\cite{pimc,Fasolino2007,los2009scaling,roldan2011suppression,wei2014graphene}. 
Even if this contradicts our results since such atomistic calculations respect in principle rotational symmetry, an uncontrollable strain in the numerical simulations as small as $\delta a=10^{-5}$ is enough to lower the exponent from $-4$ to $-3.2$ in the long wave-length limit (see Methods). Considering that the ZA mode with $q\sim 0.01\AA^{-1}$ requires about 1 nanosecond to perform one period, very long simulation times are required to describe a thermodynamically flat phase of graphene, and, thus, these Monte Carlo and MD numerical simulations may also be affected by non-ergodic conditions, affecting the determination of the height correlation function in the long wavelength limit. 
On the contrary, in our SCHA simulations the centroids are always in the plane.

In conclusion, we show that anharmonic effects are crucial to make sound propagate in graphene despite its out-of-plane acoustic mode has a quadratic dispersion as imposed by symmetry. Contrary to the previously assumed behavior, we determine that the bending rigidity of graphene does not diverge in the long wavelength limit and that the amplitude of the ripples are not suppressed by phonon-phonon interactions. These conclusions are universal and can be extrapolated to any strictly 2D material or membrane. 

%\vspace{0.5cm}

%\bibliography{bibliography}% Produces the bibliography via BibTeX.

\clearpage

\setcounter{figure}{0}
\renewcommand{\figurename}{Extended Data Figure}

\section*{Methods}

\textbf{Anharmonic theory: SCHA}.
We study the lattice dynamics of graphene in the Born-Oppenheimer (BO) approximation, thus we consider the quantum 
Hamiltonian for the atoms defined by the BO potential energy $V(\mathbf{R})$. With $\mathbf{R}$ we are denoting in component-free notation the quantity $R^{s\alpha}(\mathbf{l})$, which is a collective coordinate that completely 
specifies the atomic configuration of the crystal. The index $\alpha$ denotes the Cartesian direction, $s$ labels 
the atom within the unit cell and $\mathbf{l}$ indicates the three dimensional lattice vector. In what follows we will also use a single compact index $a=(\alpha,s,\mathbf{l})$ to indicate Cartesian index $\alpha$, atom $s$ index 
and lattice vector $\mathbf{l}$. Moreover, in general, we will use bold letters to indicate also other quantities 
in component-free notation. \\

In order to take into account quantum effects and anharmonicity at a non-perturbative level, we use the 
Self-Consistent Harmonic Approximation~\cite{errea2014anharmonic,bianco2017second,monacelli2018pressure} (SCHA). For a given temperature $T$, the method 
allows to find an approximation for $F(\boldsymbol{\mathcal{R}})$, the free energy of the crystal as a function 
of the average atomic positions $\mathcal{R}^{a}$ (the centroids). For a given centroid 
$\boldsymbol{\mathcal{R}}$, the SCHA free energy is obtained through an auxiliary quadratic Hamiltonian, the SCHA 
Hamiltonian $\mathcal{H}_{\boldsymbol{\mathcal{R}}\boldsymbol{\Phi}}$, by variationally minimizing the free energy with respect to 
the SCHA centroids and auxiliary force-constants $\boldsymbol{\Phi}$. The 
free energy Hessian, or the physical phonons in the static approach, can be computed by using the analytic formula 
(in component-free notation)
\begin{equation}\label{formula}
 \frac{\partial^{2}F}{\partial\boldsymbol{\mathcal{R}}\partial\boldsymbol{\mathcal{R}}}=\mathbf{\Phi} + \overset{(3)}{\mathbf{\Phi}}\mathbf{
 \Lambda}(0)[\mathbf{1}-\overset{(4)}{\mathbf{\Phi}}\mathbf{\Lambda}(0)]^{-1}\overset{(3)}{\mathbf{\Phi}},
\end{equation}
with
\begin{equation}
\begin{split}
% \mathbf{\Phi}=&\left\langle\frac{\partial^{2}V}{\partial\mathbf{R}\partial\mathbf{R}}\right\rangle_{\rho_{\boldsymbol{\mathcal{R}}\boldsymbol{\Phi}}}, \\ 
 \overset{(3)}{\mathbf{\Phi}}=\left\langle\frac{\partial^{3}V}{\partial\mathbf{R}\partial\mathbf{R}\partial{\mathbf{R}}}\right\rangle_{
 \rho_{\boldsymbol{\mathcal{R}}\boldsymbol{\Phi}}}& \; \overset{(4)}{\mathbf{\Phi}}=\left\langle\frac{\partial^{4}V}{\partial\mathbf{R}\partial\mathbf{
 R}\partial\mathbf{R}\partial\mathbf{R}}\right\rangle_{\rho_{\boldsymbol{\mathcal{R}}\boldsymbol{\Phi}}}, 
 \end{split}
\end{equation}
where the averages are with respect to the density matrix of the SCHA 
Hamiltonian $\mathcal{H}_{\boldsymbol{\mathcal{R}}\boldsymbol{\Phi}}$, i.e. $\rho_{\boldsymbol{\mathcal{R}}\boldsymbol{\Phi}}=e^{-\beta\mathcal{H}_{\boldsymbol{\mathcal{R}}\boldsymbol{\Phi}}}/tr[e^{-\beta\mathcal{H}_{\boldsymbol{\mathcal{R}}\boldsymbol{\Phi}}}]$, and $\beta=(K_{B}T)^{-1}$ where $K_{B}$ is the Boltzmann constant. In Eq. \eqref{formula} the value $z=0$ of the 4th-order 
tensor $\mathbf{\Lambda}(z)$ is used. For a generic complex number $z$ it is defined, in components, by
\begin{multline}
 \mathbf{\Lambda}^{abcd}(z)=-\frac{1}{2}\sum_{\mu\nu}\tilde{F}(z,\omega_{\mu},\omega_{\nu})\times \\ \times   
 \sqrt{\frac{\hbar}{2M_{a}
 \omega_{\mu}}}e_{\mu}^{a}\sqrt{\frac{\hbar}{2M_{b}\omega_{\nu}}}e_{\nu}^{b}\sqrt{\frac{\hbar}{2M_{c}\omega_{\mu}}}e_{\mu}^{
 c}\sqrt{\frac{\hbar}{2M_{d}\omega_{\nu}}}e_{\nu}^{d},
\end{multline}
with $M_{a}$ the mass of the atom $a$, $\omega_{\mu}^{2}$ (the \emph{auxiliary} phonons) the eigenvalues and $e_{\mu}^{a}$ the
eigenvectors of $D_{ab}^{(S)}=
\Phi_{ab}/\sqrt{M_{a}M_{b}}$, respectively, and 
\begin{multline}
\label{raffaello-function}
 \tilde{F}(z,\omega_{\mu},\omega_{\nu})=\frac{2}{\hbar} \Bigg[ \frac{(\omega_{\mu}+\omega_{\nu})[1+n_{B}(\tilde{
 \Omega}_{\mu})+n_{B}(\omega_{\nu})]}{(\omega_{\mu}+\omega_{\nu})^{2}-z^{2}} \\ -\frac{(\omega_{\mu}-\tilde{
 \Omega}_{\nu})[n_{B}(\omega_{\mu})-n_{B}(\omega_{\nu})]}{(\omega_{\mu}-\omega_{\nu})^{2}-z^{2}} \Bigg]
\end{multline}
where $n_{B}(\omega)=1/(e^{\beta\hbar\omega}-1)$ is the bosonic occupation number. \\ 

\begin{figure*}[ht]
\includegraphics[width=0.49\linewidth]{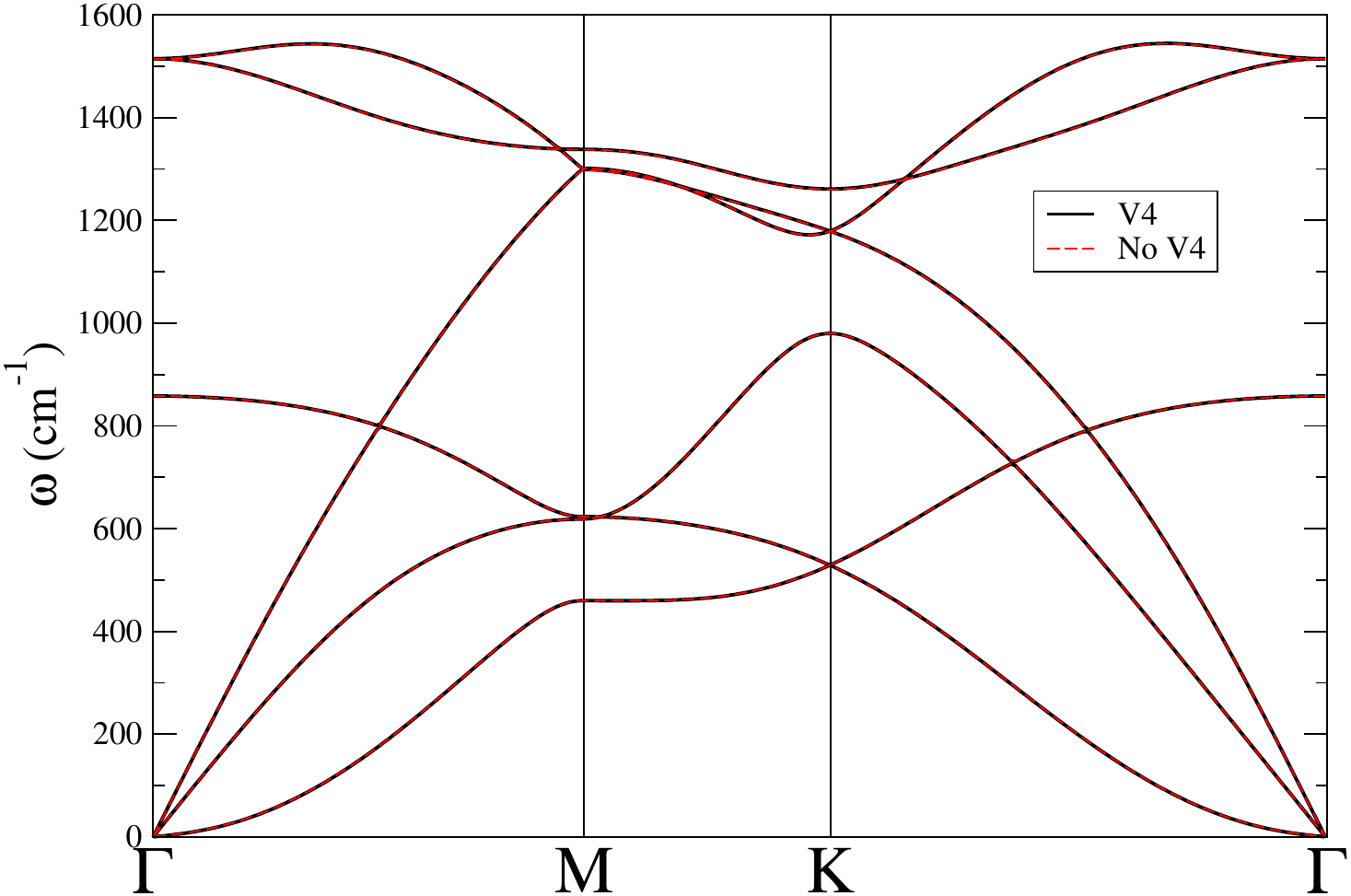}
\includegraphics[width=0.49\linewidth]{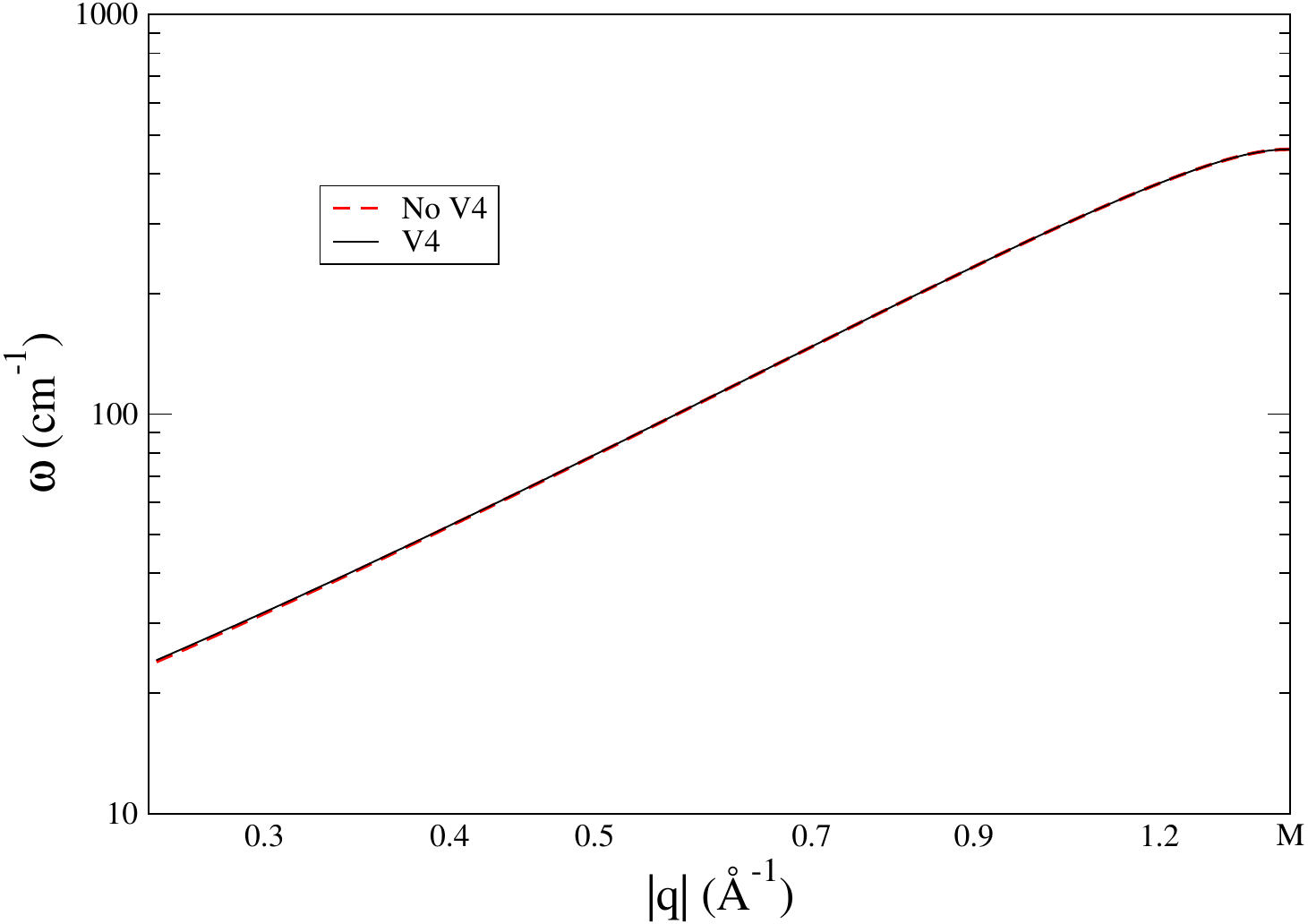}
\caption{Physcial phonons in the static approach with the atomistic potential at $500$ K including and neglecting $\overset{(4)}{\mathbf{\Phi}}$ in Eq. \eqref{scha-se}. The right panel 
only includes the ZA modes and it is in logarithmic scale. The calculation is done in a $6\times6$ supercell.}
\label{checkv4-atomistic}
\end{figure*}

As shown in Refs. ~\cite{bianco2017second,monacelli2021time,lihm2021gaussian}, in the SCHA the  Green function $\mathbf{G}(i\Omega_n)$for the correlation of 
variable $\sqrt{M_{a}}(R^{a}-\mathcal{R}^{a})$ in the frequency domain ($\Omega_n$ is a Matsubara frequency) is given as
\begin{equation} \label{green}
 \mathbf{G}^{-1}(i\Omega_n)=(i\Omega_n)^{2}\mathbf{1}-\mathbf{M}^{-\frac{1}{2}}\mathbf{\Phi}\mathbf{M}^{-\frac{1}{2}}- \mathbf{\Pi}(i\Omega_n),
\end{equation}
where $\mathbf{G}^{-1}(0)=-\mathbf{D}^{(F)}$, $D_{ab}^{(F)}=\frac{1}{\sqrt{M_{a}M_{b}}}\frac{\partial^{2}F}{\partial\mathbf{\mathcal{R}}_{a}
\partial\mathbf{\mathcal{R}}_{b}}$, and $\boldsymbol{\Pi}(z)$ is the SCHA self-energy, given by
\begin{equation}
 \boldsymbol{\Pi}(i\Omega_n)=\mathbf{M}^{-\frac{1}{2}}\overset{(3)}{\mathbf{\Phi}}\mathbf{\Lambda}(i\Omega_n)[\mathbf{1}-\overset{(4)}{\mathbf{\Phi}}\mathbf{\Lambda}(i\Omega_n)]^{-1}
 \overset{(3)}{\mathbf{\Phi}}\mathbf{M}^{-\frac{1}{2}},
 \label{scha-se}
\end{equation}
where $M_{ab}=\delta_{ab}M_{a}$ is the mass matrix. For the applications considered in the present paper, the 
static term $\overset{(4)}{\mathbf{\Phi}}\mathbf{\Lambda}(0)$ is negligible  with respect to the identity 
matrix (see Extended Data Fig. \ref{checkv4-atomistic}). Extending this approximation to the dynamical case reduces the SCHA self-energy to the so called bubble 
self-energy, namely
\begin{equation}
 \mathbf{\Pi}\approx\mathbf{\Pi}^{(B)}(i\Omega_n)=\mathbf{M}^{-\frac{1}{2}}\overset{(3)}{\mathbf{\Phi}}\mathbf{\Lambda}(i\Omega_n)\overset{(3)}{\mathbf{
 \Phi}}\mathbf{M}^{-\frac{1}{2}}.
\end{equation}
We then neglect the mixing between different phonon modes and assume that $\mathbf{\Pi}(i\Omega_n)$ is diagonal in the 
basis of the eigenvectors $e_{\mu}^{a}(\mathbf{q})$ of $\Phi_{ab}(\mathbf{q})/\sqrt{M_{a}M_{b}}$ where 
$\Phi_{ab}(\mathbf{q})$ is the Fourier transform of the real space $\boldsymbol{\Phi}$ (now $a$ and $b$ represent 
atoms in the unit cell and Cartesian indices). We then define 
\begin{equation}
 \Pi_{\mu}(\mathbf{q},i\Omega_n)=\sum_{a,b}e_{\mu}^{a}(-\mathbf{q})\Pi_{ab}(\mathbf{q},i\Omega_n)e_{\mu}^{b}
 (\mathbf{q}).
\end{equation}
\\

\begin{figure*}[ht]
\includegraphics[width=0.49\linewidth]{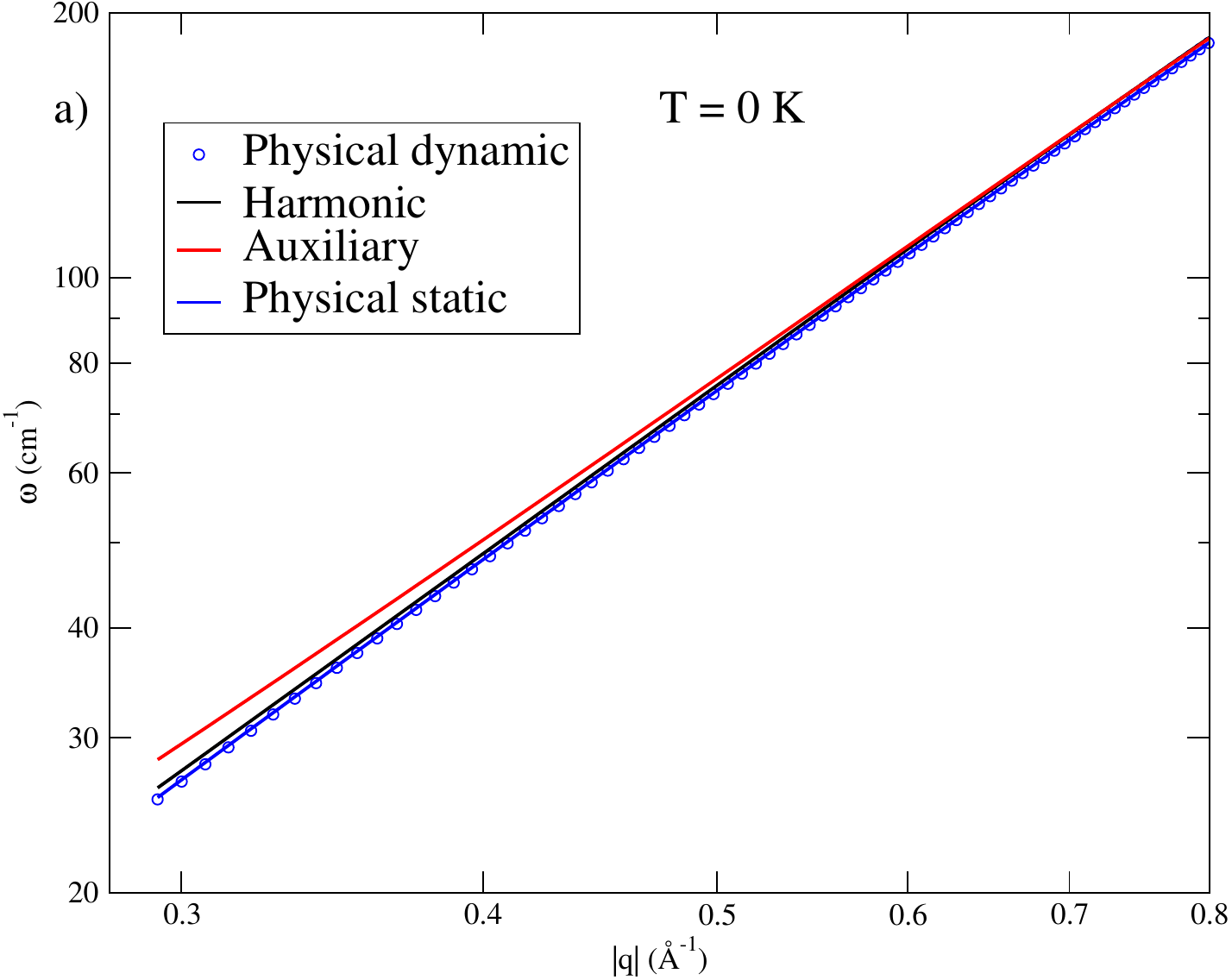}
\includegraphics[width=0.49\linewidth]{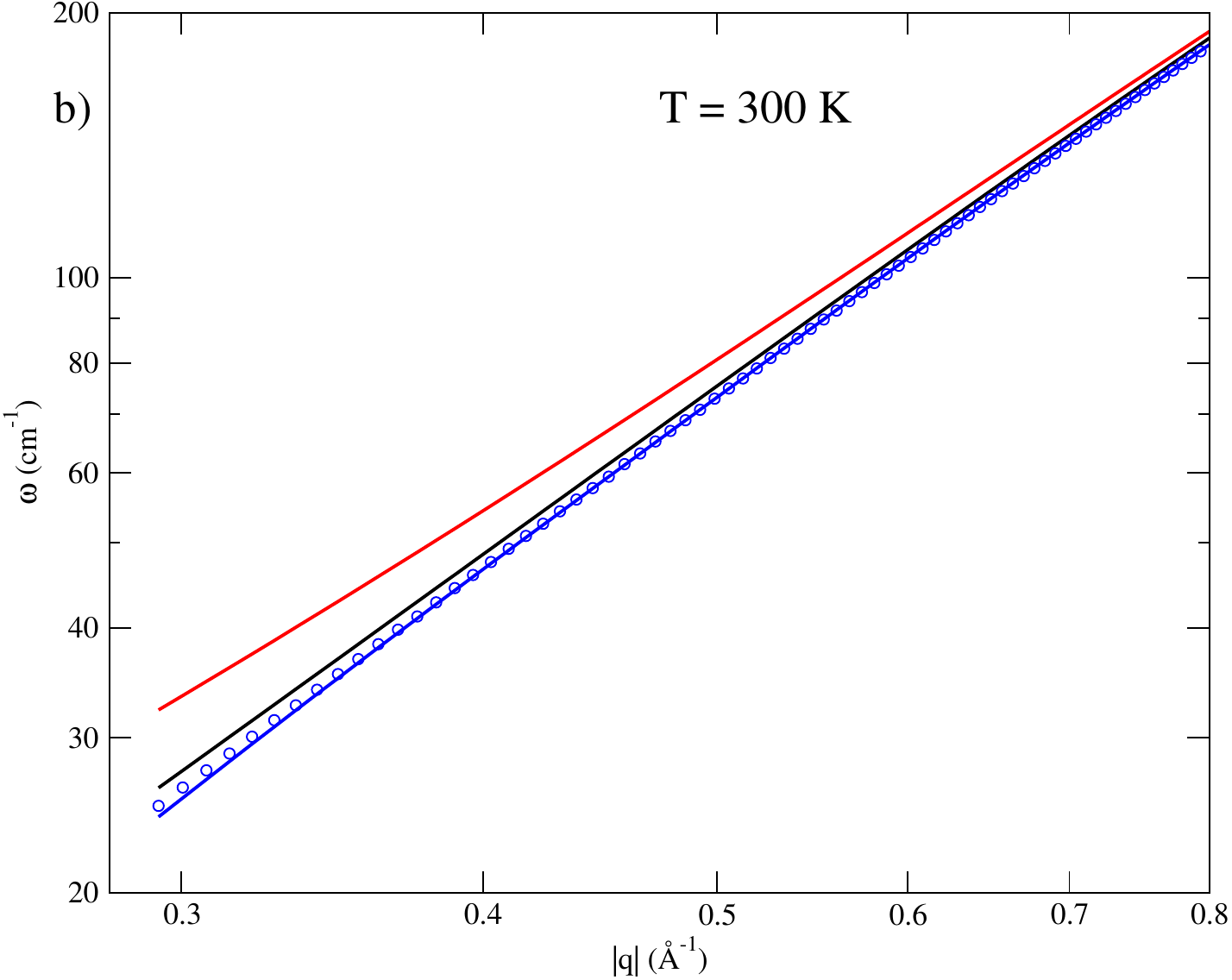}
\caption{Harmonic, and SCHA auxiliary and physical phonons (static and dynamic) calculated at 0 K (a) and 300 K (b) with the atomistic potential for the ZA mode.}
\label{static-dynamic-ed}
\end{figure*}

In studying the response of a lattice to inelastic scattering experiments we need the one-phonon spectral function. By using Eq. \eqref{green} for $\mathbf{G}(i\Omega_n)$ we can calculate the cross-section 
$\sigma(\omega)=-\omega TrIm\mathbf{G}(\omega+i0^{+})/\pi$, whose peaks signal the presence of collective 
vibrational excitations (physical phonons in the dynamic approach) having certain energies. Again, we take advantage of the lattice periodicity and we Fourier 
transform the interesting quantities with respect to the lattice indices. In particular, we consider the Fourier 
transform of the SCHA self-energy, $\Pi_{ab}(\mathbf{q},i\Omega_n)$. Neglecting the mixing between different modes, the 
cross section is then given by
\begin{equation}
\small
 \sigma(\mathbf{q},\omega)=\frac{1}{\pi}\sum_{\mu}\frac{-\omega Im\Pi_{\mu}(\mathbf{q},\omega)}{(\omega^{2}-\omega_{\mu}^{2}(\mathbf{q})-
 Re\Pi_{\mu}(\mathbf{q},\omega))^{2}+(Im\Pi_{\mu}(\mathbf{q},\omega))^{2}}.
\end{equation}
If we neglect the frequency dependence of the phonon self-energy, we get the weakly anharmonic limit of the cross section, which is going to be a sum of Lorentzian functions. These Lorentzians are well defined physical phonons in the dynamical approach. The phonon frequencies squared, $\Theta_{\mu}^{2}(\mathbf{q})$, corrected by the bubble self-energy are obtained as 
\begin{equation}
\label{dynamic-phonon}
\Theta_{\mu}^{2}(\mathbf{q})=\omega_{\mu}^{2}(\mathbf{q})+Re\Pi_{\mu}(\mathbf{q},\omega_{\mu}(\mathbf{q})), 
\end{equation}
where $\omega_{\mu}^{2}(\mathbf{q})$ are the eigenvalues of the Fourier transform of $\mathbf{D}^{(S)}$. The linewidth of the phonons in Eq. \eqref{dynamic-phonon} is proportional to 
$Im\Pi_{\mu} ( \mathbf{q}, \omega_{\mu}(\mathbf{q}) )$. The centers of these peaks are the ones supposed to be measured in inelastic experiments. By calculating  $\Omega_{\mu}^{2}(\mathbf{q}) = \omega_{\mu}^{2}(\mathbf{q})+Re\Pi_{\mu}(\mathbf{q},0)$ the static limit in Eq. \eqref{formula} is 
recovered, i.e., the eigenvalues of the free energy Hessian. 
We show in Extended Data Fig. \ref{static-dynamic-ed} that the dynamic effects are negligible in the ZA modes, meaning that the static approximation and the phonons coming from the free energy Hessian are a good approximation for the physical phonons.\\ 

\begin{figure*}[h!]
\centering
\includegraphics[width=0.69\linewidth]{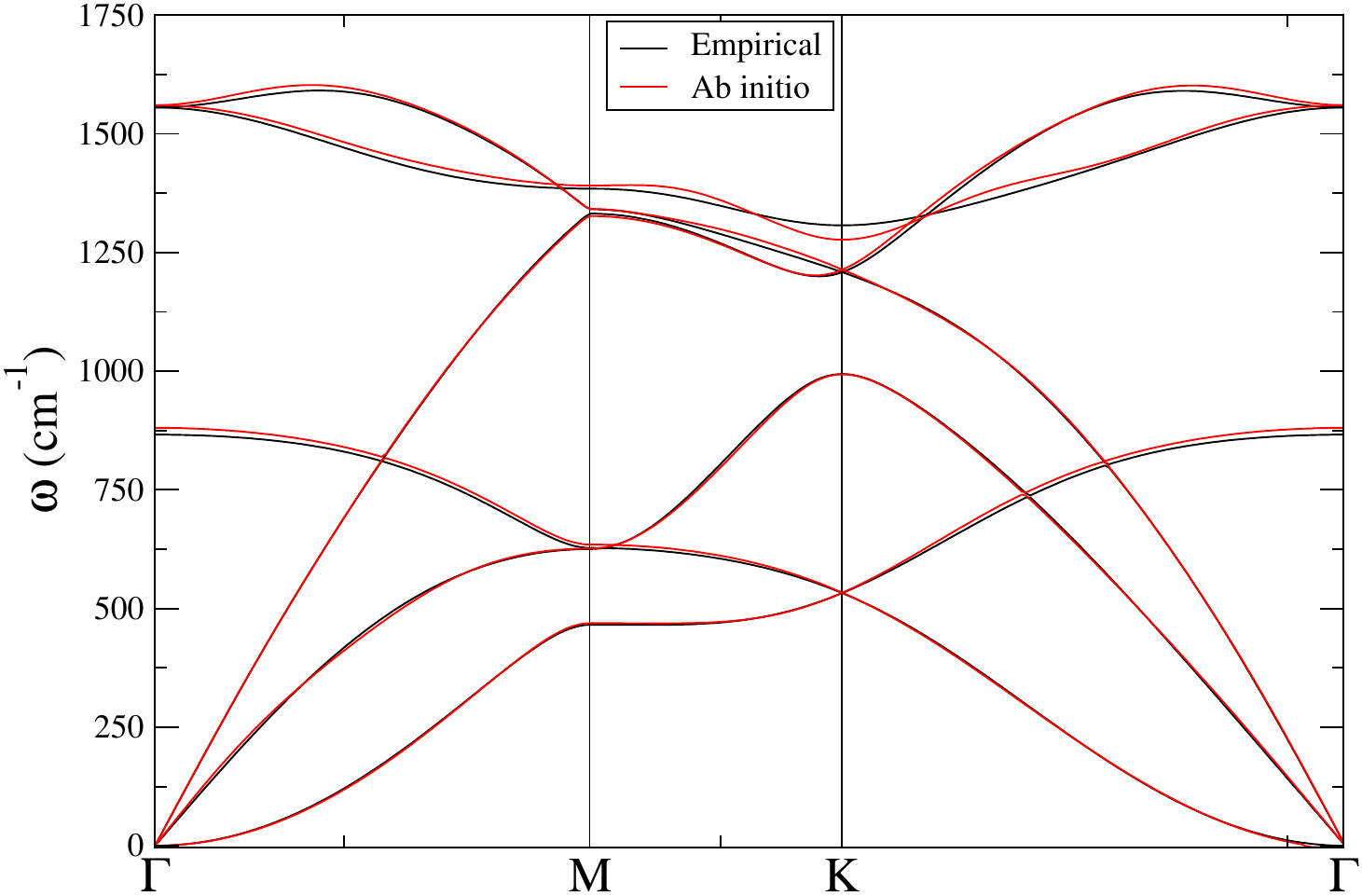}
\caption{Harmonic phonon spectrum of graphene calculated with the machine learning empirical potential and $ab$ $initio$. The calculations are done in a $6\times6$ supercell.}
\label{benchmark-spectrum-ed}
\end{figure*}

\begin{figure*}[h!]
\includegraphics[width=0.49\linewidth]{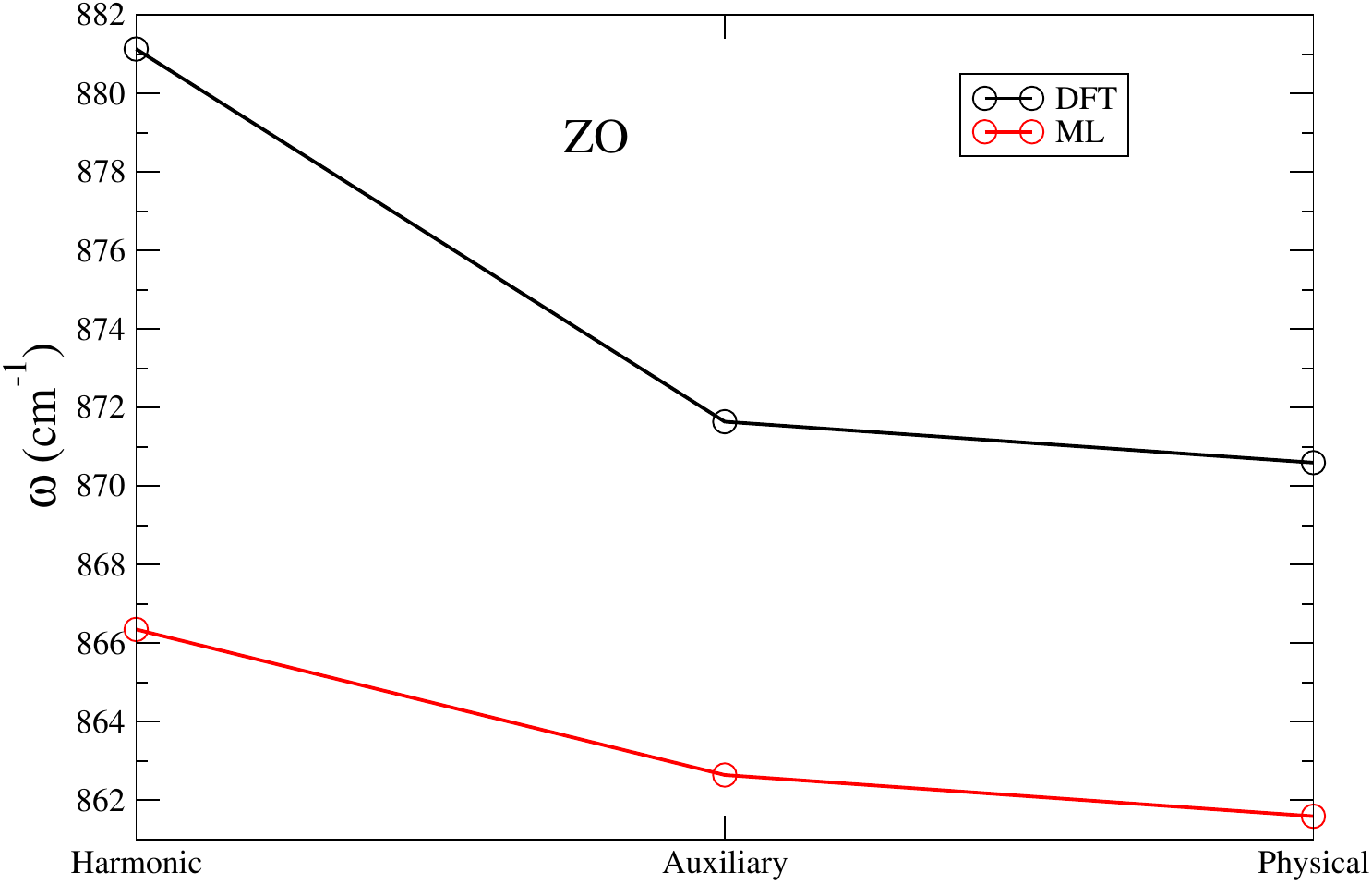}
\includegraphics[width=0.49\linewidth]{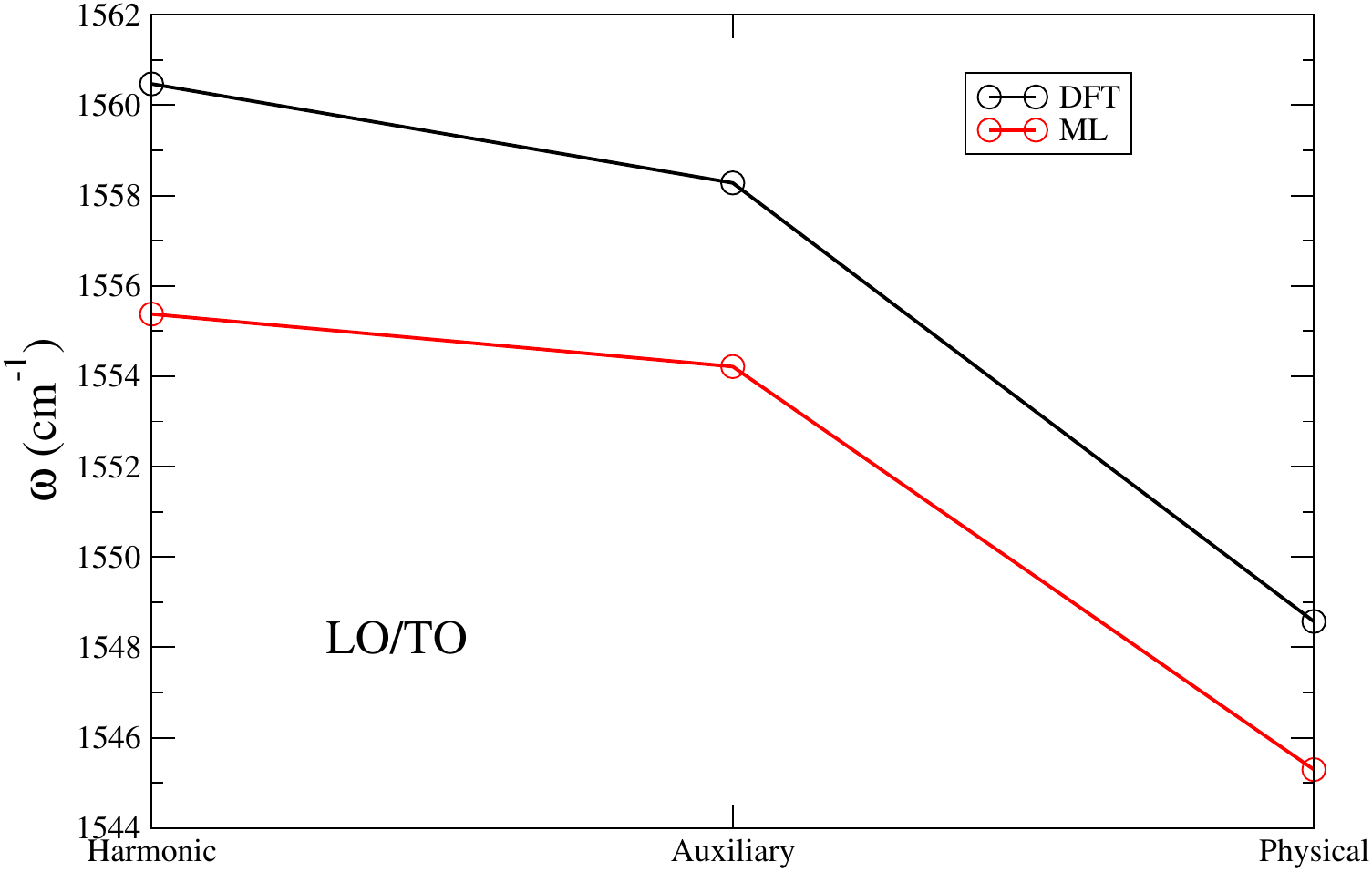}
\caption{Harmonic, and SCHA auxiliary and physical frequencies (static) using the DFT and machine learning (ML) forces. The 
left panel shows the in-plane optical frequency at the $\Gamma$ point and the right panel the out-of-plane one.}
\label{benchmark-ed}
\end{figure*}

\textbf{Empirical potential benchmark and calculation parameters of the atomistic calculations}.
For calculating the forces needed in the atomistic SCHA minimization~\cite{errea2014anharmonic} we have used an empirical 
potential trained with machine learning and density functional theory (DFT) forces. The details about the machine 
learning training are explained in Ref. ~\cite{rowe2018development}. Here we have benchmarked the ability of 
the potential to account for anharmonic effects. For that purpose we have applied the SCHA method by using 
DFT and empirical forces in a $2\times2$ supercell and we have checked the anharmonic effects in the optical 
modes at the $\Gamma$ point. The machine learning potential is trained with the exchange-correlation in 
Ref. ~\cite{dion2004van} and for the DFT calculations we have applied a PBE~\cite{perdew1996generalized} 
ultrasoft pseudopotential~\cite{vanderbilt1990soft} with Van der Walls corrections~\cite{barone2009role}. The 
results are shown in Extended Data Figs. \ref{benchmark-spectrum-ed} and \ref{benchmark-ed}.
As we can see in Extended Data Fig. \ref{benchmark-spectrum-ed}, the two potentials provide very similar harmonic phonons. Due to the different exchange correlation functional there is a slight offset in Extended Data Fig. \ref{benchmark-ed}, however, the anharmonic lineshifts  are very well captured within the empirical potential. \\
For the self-consistent DFT calculations used in the benchmark we have used a plane wave cutoff of $70$ Ry and a $700$ Ry cutoff for the 
density. For the Brillouin zone integration we have used a Monkhorst pack grid~\cite{monkhorst1976special} of 
$32\times32$ points with a Gaussian smearing of $0.02$ Ry. 

The atomistic calculations of the linewidth in the main text  have been performed 
with  a grid of $400\times400$ momentum points for the bubble self-energy, with a Gaussian smearing ($\delta$) of $1$ cm$^{-1}$. For the stress calculation in order 
to account for the thermal expansion we have used a $10\times10$ supercell. We have used the same supercell for 
the SCHA auxiliary and physical frequency calculations in the atomistic case. For the linewidth calculations we have used $\overset{(3)}{\mathbf{
 \Phi}}$  calculated in 
a $3\times3$ supercell and fourier interpolate it. We have tested all the calculations with denser grids and bigger supercells.\\

\begin{figure*}[h!]
\begin{center}
\includegraphics[width=0.49\linewidth]{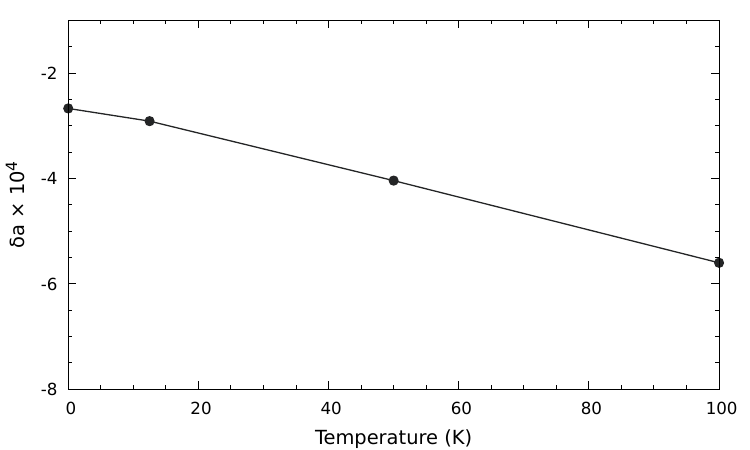}
\end{center}
\caption{$\delta a$ as a function of temperature in the membrane model.}
\label{ctemembrane}
\end{figure*}

\textbf{SCHA applied to the continuum membrane Hamiltonian}.
The general rotationally invariant potential for a membrane can be written as follows
\begin{equation}
 \label{full-potential}
 V=\frac{1}{2}\int_{\Omega}{d^{2}x\left(\kappa(\partial^{2}h)^{2}+\sum_{n\geq 2}u_{i_{1}j_{1}}\dots u_{i_{n}j_{n}}C_{i_{1}j_{1}\dots i_{n}j_{n}}^{(2n)}\right)},
\end{equation}
where $\Omega$ is the area of the membrane in equilibrium, $\kappa$ is the bending rigidity, $h$ is the 
out-of-plane component of the displacement field and the rotationally invariant strain tensor $u_{ij}$ is defined 
using the in-plane displacement field $u_{i}$
\begin{equation}
 \label{strain-tensor}
 u_{ij}=\frac{1}{2}(\partial_{i}u_{j}+\partial_{j}u_{i}+\partial_{i}\boldsymbol{u}\cdot\partial_{j}\boldsymbol{u}+\partial_{i}h\partial_{j}h).
\end{equation}
$C^{(2n)}_{i_{1}j_{1}\dots i_{n}j_{n}}$ is the generic elastic tensor of rank $2n$. In the previous expression the 
subscripts label the 2D coordinates $x,y$ and the sum over indices is assumed. The second-order expansion of 
Eq. \eqref{full-potential} with respect to the phonon fields is given by
\begin{equation}
 \label{rank2-potential-strain}
 V=\frac{1}{2}\int_{\Omega}{d^{2}x\left(\kappa(\partial^{2}h)^{2}+C^{(4)}_{ijkl}u_{ij}u_{il}\right)},
\end{equation}
with $C^{(4)}_{ijkl}=\lambda\delta_{ij}\delta_{kl}+\mu(\delta_{ik}\delta_{jl}+\delta_{il}\delta_{jk})$. By using 
equation \eqref{strain-tensor} and $C_{ijkl}^{(4)}=C^{ijkl}$, equation \eqref{rank2-potential-strain} can be rewritten as
\begin{multline}
 \label{rank2-potential-disp}
 V=\frac{1}{2}\int_{\Omega}d^{2}x[\kappa(\partial^{2}h)^{2}+C^{ijkl}\partial_{i}u_{j}\partial_{k}u_{l}+C^{ijkl}\partial_{i}u_{j}\partial_{k}h\partial_{l}h+ \\ +\frac{C^{ijkl}}{4}\partial_{i}h\partial_{j}h\partial_{k}h\partial_{l}h+\frac{C^{ijkl}}{2}\partial_{i}\boldsymbol{u}\cdot\partial_{j}\boldsymbol{u}\partial_{k}h\partial_{l}h+ \\ + C^{ijkl}\partial_{i}u_{j}\partial_{k}\boldsymbol{u}\cdot\partial_{l}\boldsymbol{u}+
\frac{C^{ijkl}}{4}\partial_{i}\boldsymbol{u}\cdot\partial_{j}\boldsymbol{u}\partial_{k}\boldsymbol{u}\cdot\partial_{l}\boldsymbol{u}].
\end{multline}
If we allow the lattice spacing $a$ to be a variable, we can vary it by simply shifting the derivatives of the in-plane displacements according to $\partial_{i}u_{j}\rightarrow\partial_{i}u_{j}+\delta^{ij}\delta a$, where
$\delta a=(a-a_{0})/a_{0}$. Then, by taking into account periodic boundary conditions,  $\int_{\Omega}{d^{2}x\partial_{i}u_{j}}=0$, and we can rewrite the potential as
\begin{multline}
 \small
 V\rightarrow V+2\Omega(1+\delta a)(\lambda+\mu)\delta a^{2}+ \\+(1+\frac{\delta a}{2})\delta a(\lambda+\mu)\int_{\Omega}{d^{2}x\partial_{k}h\partial_{k}h}+\\
 +\frac{\delta a}{2}\int_{\Omega}{d^{2}xC^{ijkl}\partial_{i}u_{j}\partial_{k}h\partial_{
 l}h}+\\+(1+\frac{\delta a}{2})\delta a\int_{\Omega}{d^{2}xC^{ijkl}\partial_{i}u_{j}\partial_{k}u_{l}}+\\+(1+\frac{\delta a}{2})\delta a(\lambda+\mu)\int_{\Omega}{d^{2}x\partial_{k}\boldsymbol{
 u}\cdot\partial_{k}\boldsymbol{u}}+\frac{\delta a^{4}\Omega}{2}(\lambda+\mu)+\\+\frac{\delta a}{4}\int_{\Omega}{d^{2}xC^{ijkl}[\partial_{i}\boldsymbol{u}\cdot\partial_{j}\boldsymbol{u}\partial_{k}u_{l}+\partial_{
 i}u_{j}\partial_{k}\boldsymbol{u}\cdot\partial_{l}\boldsymbol{u}]}.  
 \label{eq:ptential_da}
\end{multline}
The displacement fields $\textbf{u}(\textbf{x}), h(\textbf{x})$ can be expanded in the following plane wave basis set:
\begin{equation} 
	 \textbf{u}(\textbf{x})= \frac{1}{\sqrt{\Omega}} \sum_{\textbf{q}} \textbf{u}(\textbf{q}) e^{i\textbf{q} \cdot \textbf{x}},
\end{equation}
\begin{equation} 
	  h(\textbf{x}) =  \frac{1}{\sqrt{\Omega}} \sum_{\textbf{q}} h(\textbf{q}) e^{i\textbf{q} \cdot \textbf{x}},
\end{equation}
where \textbf{q} are discrete wavevectors determined by periodic boundary conditions and $\textbf{u}(\textbf{q}), h(\textbf{q})$ the corresponding Fourier transforms, which are defined according to
\begin{equation} 
	 \textbf{u}(\textbf{q})= \frac{1}{\sqrt{\Omega}} \int_{\Omega} d^2x \; \textbf{u}(\textbf{x}) e^{-i\textbf{q} \cdot \textbf{x}},
\end{equation}
\begin{equation} 
	 h(\textbf{q})= \frac{1}{\sqrt{\Omega}} \int_{\Omega} d^2x \; h(\textbf{x}) e^{-i\textbf{q} \cdot \textbf{x}}.
\end{equation}
Then, the SCHA free energy can be written as (we use $\hbar=k_{B}=1$):
\begin{multline}
\label{scha-f}
 \mathcal{F}(\mathcal{V})=F_{\mathcal{V}}+2\Omega(1+\delta a+\frac{\delta a^{2}}{4})(\lambda+\mu)\delta a^{2}+\\ 
 +\frac{1}{2}\sum_{\boldsymbol{q}}\Biggl\{g[\omega_{SCHA}^{(h)}(\boldsymbol{q})]\kappa|\boldsymbol{q}|^{4}+\\+
 \{(\lambda+2\mu)
 g[\omega_{SCHA}^{(LA)}(\boldsymbol{q})]
 +\mu g[\omega_{SCHA}^{(TA)}(\boldsymbol{q})]\}|\boldsymbol{q}|^{2}+\\
 +\frac{\lambda+2\mu}{4\Omega}\sum_{\boldsymbol{k}}g[\omega_{SCHA}^{(h)}(\boldsymbol{q})]g[\omega_{SCHA}^{(h)} (\boldsymbol{k})][|\boldsymbol{q}|^{2}|\boldsymbol{k}|^{2}+2(\boldsymbol{q}\cdot\boldsymbol{k})^{2}]+ \\ 
 +\frac{1}{2\Omega}\sum_{\boldsymbol{k}}g[\omega_{SCHA}^{(h)}(\boldsymbol{k})]\{g[\omega_{SCHA}^{(LA)}(\boldsymbol{q})]+g[\omega_{SCHA}^{(TA)}(
 \boldsymbol{q})]\}\times \\
 \times[\lambda|\boldsymbol{q}|^{2}|\boldsymbol{k}|^{2}+2\mu(\boldsymbol{q}\cdot\boldsymbol{k})^{2}]+ \\ 
 +2(1+\frac{\delta a}{2})\delta a(\lambda+\mu)g[\omega_{SCHA}^{(h)}(\boldsymbol{q})]|\boldsymbol{q}|^{2} + \\ 
 +2(1+\frac{ \delta a}{2})\delta a\{(\lambda+2\mu)g[\omega_{SCHA}^{(LA)}(\boldsymbol{q})]+\mu g[\omega_{SCHA}^{(TA)}(\boldsymbol{q})]\}|\boldsymbol{q}|^{2}+ \\
 +2(1+\frac{\delta a}{2})\delta a(\lambda+\mu)\{g[\omega_{SCHA}^{(LA)}(\boldsymbol{
 q})]+g[\omega_{SCHA}^{(TA)}(\boldsymbol{q})]\}|\boldsymbol{q}|^{2}+\\
 + 
 \frac{1}{4\Omega}\sum_{\boldsymbol{k}}\bigg[4g[\omega_{SCHA}^{(LA)}(
 \boldsymbol{q})]g[\omega_{SCHA}^{(TA)}(\boldsymbol{k})]\times \\ 
 \times[\lambda(\boldsymbol{q}\cdot\boldsymbol{k})^{2}+\mu|\boldsymbol{q}|^{2}|\boldsymbol{k}|^{2}+\mu(\boldsymbol{q}\cdot\boldsymbol{k})^{2}](\hat{\boldsymbol{q}_{\perp}}\cdot\hat{
 \boldsymbol{k}})+ \\ + 2g[\omega_{SCHA}^{(LA)}(\boldsymbol{q})]g[\omega_{SCHA}^{(TA)}(\boldsymbol{k})][\lambda|\boldsymbol{q}|^{2}|\boldsymbol{k}|^{2}+2\mu(\boldsymbol{q}\cdot\boldsymbol{k})^{2}]+ \\ + (g[\omega_{SCHA}^{(LA)}(
 \boldsymbol{q})]g[\omega_{SCHA}^{(LA)}(\boldsymbol{k})]+g[\omega_{SCHA}^{(TA)}(\boldsymbol{q})]g[\omega_{SCHA}^{(TA)}(\boldsymbol{k})])\times \\ 
 \times[\lambda|\boldsymbol{q}|^{2}|\boldsymbol{k}|^{2}+2\mu(\boldsymbol{q}\cdot\boldsymbol{k})^{2}]+ 
 \\2(g[\omega_{SCHA}^{(LA)}(\boldsymbol{q})]g[\omega_{SCHA}^{(LA)}(\boldsymbol{k})]) \times \\
 \times[\lambda(\boldsymbol{q}\cdot\boldsymbol{k})^{2}+\mu|\boldsymbol{q}|^{2}|\boldsymbol{k}|^{2}+\mu(\boldsymbol{q}\cdot\boldsymbol{k})^{2}](\hat{
 \boldsymbol{q}}\cdot\hat{\boldsymbol{k}}) + \\ + 2(g[\omega_{SCHA}^{(TA)}(\boldsymbol{q})]g[\omega_{SCHA}^{(TA)}(\boldsymbol{k})])\times \\ 
 \times[\lambda(\boldsymbol{q}\cdot\boldsymbol{k})^{2}+\mu|\boldsymbol{q}|^{2}|\boldsymbol{k}|^{2}+\mu(\boldsymbol{q}\cdot\boldsymbol{k})^{2}](\hat{\boldsymbol{q}_{\perp}}\cdot\hat{\boldsymbol{k}_{\perp}}) \bigg]+ \\%
% -g[\omega_{SCHA}^{(h)}(\boldsymbol{q})]\Phi_{SCHA}^{(h)}(\boldsymbol{q}) ESTO YA VA EN EL BATUKARI
-\sum_{\alpha}g[\omega_{SCHA}^{(\alpha)}(\boldsymbol{q})]\Phi_{SCHA}^{(\alpha)}(\boldsymbol{q}) \Biggr\},
\end{multline}
where $g(\omega)=coth((\omega/2T))/(2\rho\omega)$ and $\omega_{SCHA}^{\alpha}(\boldsymbol{q})=\sqrt{\Phi_{SCHA}^{(\alpha)}(
\boldsymbol{q})/\rho}$ ($\alpha=h,LA,TA$) is the SCHA auxiliary frequency. $\rho$ is the mass density. In Eq. \eqref{scha-f} the in-plane displacement vector $\boldsymbol{u}(\boldsymbol{q}$) is separated into  longitudinal and transversal components $\boldsymbol{u}(\boldsymbol{q})=u_{LA}(\boldsymbol{q})\hat{\boldsymbol{q}}+u_{TA}(\boldsymbol{q})\hat{\boldsymbol{q}_{\perp}}$, $\hat{\boldsymbol{q}}_{\perp}$ being the unitary vector perpendicular to $\hat{\boldsymbol{q}}$. $F_{\mathcal{V}}$ is the harmonic free energy of the harmonic auxiliary potential $\mathcal{V}$. Now, by taking the derivative of the SCHA free energy with respect to the lattice constant and SCHA auxiliary frequencies, we arrive to the SCHA equations:
\begin{multline}
 \frac{\partial\mathcal{F}(\mathcal{V})}{\partial\delta a}=0=2\Omega(2\delta a+3\delta a^{2}+\delta a^{3})(\lambda+
\mu)+\\
+\frac{1}{2}\sum_{\boldsymbol{q}}g[\omega_{SCHA}^{(h)}(\boldsymbol{q})]2(1+\delta a)(\lambda+\mu)|\boldsymbol{
 q}|^{2}+\\+\frac{1}{2}\sum_{\boldsymbol{q}}g[\omega_{SCHA}^{(LA)}(\boldsymbol{q})][2(1+\delta a)(\lambda+2\mu)|\boldsymbol{q}|^{2}+2(1+\delta a)(\lambda+\mu)|\boldsymbol{q}|^{2}]+ \\ +\frac{1}{2}\sum_{\boldsymbol{q}}g[\Omega_{
 SCHA}^{(TA)}(\boldsymbol{q})][2(1+\delta a)\mu|\boldsymbol{q}|^{2}+2(1+\delta a/2)(\lambda+\mu)|\boldsymbol{q}|^{2}],
\end{multline}
\begin{multline}
 \Phi_{SCHA}^{(h)}(\boldsymbol{q})=\kappa|\boldsymbol{q}|^{4}+2(1+\delta a/2)\delta a(\lambda+\mu)|\boldsymbol{q}|^{2}+\\+\frac{\lambda+2\mu}{2\Omega}\sum_{\boldsymbol{k}}g[\omega_{SCHA}^{(h)}(\boldsymbol{k})][|\boldsymbol{q}|^{2}|\boldsymbol{
 k}|^{2}+2(\boldsymbol{q}\cdot\boldsymbol{k})^{2}]+ \\ +\frac{1}{2\Omega}\sum_{\boldsymbol{k}}\{g[\omega_{SCHA}^{(LA)}(\boldsymbol{k})]+g[\omega_{SCHA}^{(TA)}(\boldsymbol{k})]\}[\lambda|\boldsymbol{q}|^{2}|\boldsymbol{k}|^{2}+2\mu(
 \boldsymbol{q}\cdot\boldsymbol{k})^{2}],
\end{multline}
\begin{multline}
 \Phi_{SCHA}^{(LA)}(\boldsymbol{q})=(\lambda+2\mu)|\boldsymbol{q}|^{2}+2(1+\delta a/2)\delta a(\lambda+2\mu)|\boldsymbol{q}|^{2}+\\+2(1+\delta a/2)\delta a(\lambda+\mu)|\boldsymbol{q}|^{2}+\\+\frac{1}{2\Omega}\sum_{\boldsymbol{k}}g[\omega_{SCHA}^{
 (h)}(\boldsymbol{k})][\lambda|\boldsymbol{q}|^{2}|\boldsymbol{k}|^{2}+2\mu(\boldsymbol{q}\cdot\boldsymbol{k})^{2}]+\\+\frac{1}{4\Omega}\sum_{\boldsymbol{k}}\biggl\{4g[\omega_{SCHA}^{(TA)}(\boldsymbol{k})][\lambda(\boldsymbol{q}\cdot\boldsymbol{
 k})^{2}+\mu|\boldsymbol{q}|^{2}|\boldsymbol{k}|^{2}+\mu(\boldsymbol{q}\cdot\boldsymbol{k})^{2}](\hat{\boldsymbol{q}_{\perp}}\cdot\hat{\boldsymbol{k}})+ \\ + 2g[\omega_{SCHA}^{(TA)}(\boldsymbol{k})][\lambda|\boldsymbol{q}|^{2}|\boldsymbol{
 k}|^{2}+2\mu(\boldsymbol{q}\cdot\boldsymbol{k})^{2}]+ \\ + 2g[\omega_{SCHA}^{(LA)}(\boldsymbol{k})][\lambda|\boldsymbol{q}|^{2}|\boldsymbol{k}|^{2}+2\mu(\boldsymbol{q}\cdot\boldsymbol{k})^{2}]+ \\ + 4g[\omega_{SCHA}^{(LA)}(\boldsymbol{
 k})][\lambda(\boldsymbol{q}\cdot\boldsymbol{k})^{2}+\mu|\boldsymbol{q}|^{2}|\boldsymbol{k}|^{2}+\mu(\boldsymbol{q}\cdot\boldsymbol{k})^{2}](\hat{\boldsymbol{q}}\cdot\hat{\boldsymbol{k}})\biggr\}
\end{multline}
and,
\begin{multline}
 \Phi_{SCHA}^{(TA)}(\boldsymbol{q})=\mu|\boldsymbol{q}|^{2}+2(1+\delta a/2)\delta a\mu|\boldsymbol{q}|^{2}+\\+2(1+\delta a/2)\delta a(\lambda+\mu)|\boldsymbol{q}|^{2}+\\+\frac{1}{2\Omega}\sum_{\boldsymbol{k}}g[\omega_{SCHA}^{
 (h)}(\boldsymbol{k})][\lambda|\boldsymbol{q}|^{2}|\boldsymbol{k}|^{2}+2\mu(\boldsymbol{q}\cdot\boldsymbol{k})^{2}]+\\+\frac{1}{4\Omega}\sum_{\boldsymbol{k}}\biggl\{
 4g[\omega_{SCHA}^{(TA)}(\boldsymbol{k})][\lambda(\boldsymbol{q}\cdot\boldsymbol{k})^{2}+\mu|\boldsymbol{q}|^{2}|\boldsymbol{k}|^{2}+\mu(\boldsymbol{q}\cdot\boldsymbol{k})^{2}](\hat{\boldsymbol{q}_{\perp}}\cdot\hat{
 \boldsymbol{k}_{\perp}})+ \\ +4g[\omega_{SCHA}^{(LA)}(\boldsymbol{k})][\lambda(\boldsymbol{q}\cdot\boldsymbol{k})^{2}+\mu|\boldsymbol{q}|^{2}|\boldsymbol{k}|^{2}+\mu(\boldsymbol{q}\cdot\boldsymbol{k})^{2}](\hat{\boldsymbol{
 q}_{\perp}}\cdot\hat{\boldsymbol{k}})+ \\ + 2g[\omega_{SCHA}^{(TA)}(\boldsymbol{k})][\lambda|\boldsymbol{q}|^{2}|\boldsymbol{k}|^{2}+2\mu(\boldsymbol{q}\cdot\boldsymbol{k})^{2}]  \biggr\}.
\end{multline}
When solving this set of equations, it has been taken into account that the assumed periodic boundary conditions make the reciprocal space discrete. In order to reach wave vectors a  magnitude of order smaller than in our atomistic calculations, we have worked with a squared membrane of size $L_x=L_y=\frac{2 \pi}{0.01}\AA$. On the other side, the implicit continuity of the membrane Hamiltonian makes Fourier transforms to be non-periodic. 
Then, as displacement fields \textbf{u}(\textbf{x}) and h(\textbf{x}) are smooth functions in  real space, their discrete and non-periodic Fourier transforms \textbf{u}(\textbf{q}) and h(\textbf{q}) (and related magnitudes) are expected to decay rapidly in  reciprocal space. Therefore, we can converge our results with respect to a cut-off radius in  momentum space, defining in this way a circular grid. The value of this cut-off radius is temperature dependent, because modes with greater q values are thermally excited when increasing the temperature. We have found that with a value of R$_{cut} = 0.8$ $\AA^{-1}$ convergence is achieved for temperatures close to 0K. This radius encloses  20080 \textbf{q}-points, which yields a total of 60241 coupled equations that we have solved by applying the Newton-Raphson method~\cite{ypma1995historical}.
This model accounts for the negative thermal expansion of graphene as it can be seen in Extended Data Fig. \ref{ctemembrane}. \\

\begin{figure*}[h!]
\includegraphics[width=0.49\linewidth]{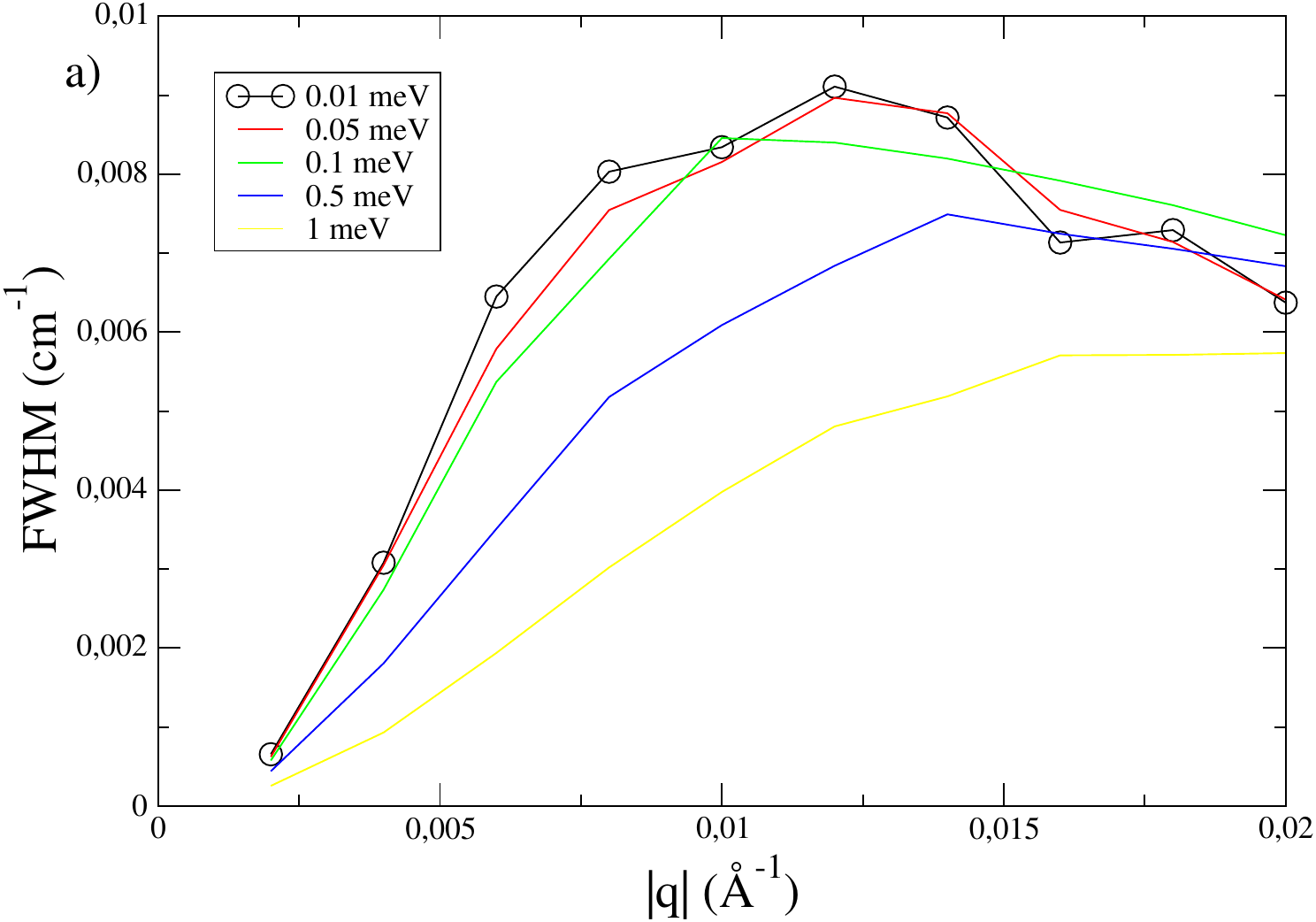}
\includegraphics[width=0.45\linewidth]{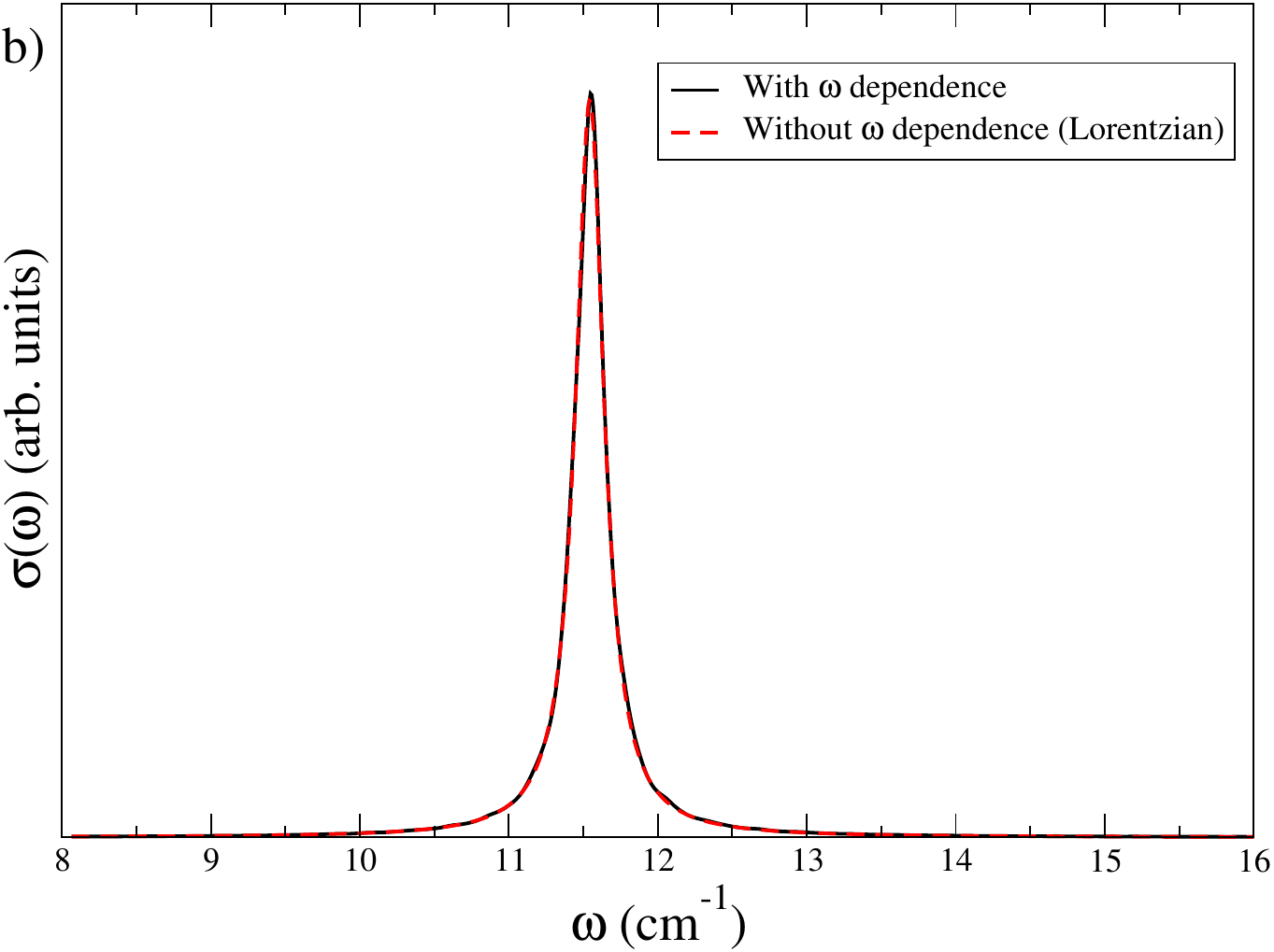}
\caption{(a) Linewidth (full width at half maximum, FWHM) contribution of the term containing the fourth-order tensor of the LA mode calculated in the membrane model at 100 K using the harmonic and SCHA auxiliar phonons. The value of the smearing is in the legend. (b) Spectral function of the LA mode with momentum 0.01 $\AA^{-1}$ with and without considering the frequency dependence of the self energy.}
\label{lw_membrane}
\end{figure*}

Regarding the second derivative of the free energy, the physical phonons in the static approach, the most general 
formula for the correction to the SCHA auxiliary phonon frequencies is
\begin{multline}
 \label{correction}
 D_{\alpha\beta}^{corr}(-\boldsymbol{q},\boldsymbol{q})=\sum_{\gamma\delta\epsilon\zeta}\sum_{\boldsymbol{p}\boldsymbol{k}}\overset{(3)}{D}{}_{\alpha\gamma\delta}(-\boldsymbol{q},\boldsymbol{p},
 \boldsymbol{q}-\boldsymbol{p}) \times \\ \times  
 [1-\overset{(4)}{D}{}_{\gamma\delta\epsilon\zeta}(-\boldsymbol{p},\boldsymbol{p}-\boldsymbol{q},\boldsymbol{k},\boldsymbol{q}-\boldsymbol{k})]^{-1}\overset{(3)}{D}{}_{
 \epsilon\zeta\beta}(-\boldsymbol{k},\boldsymbol{k}-\boldsymbol{q},\boldsymbol{q}),
\end{multline}
where the subindexes run on the normal coordinates $\alpha,\beta,\gamma,\delta,\epsilon,\zeta=h,u_{LA},u_{TA}$ and the dynamical matrices in normal coordinates are defined as
\begin{equation}
 \label{third-order}
 \overset{(3)}{D}{}_{\alpha\beta\gamma}(\boldsymbol{q},\boldsymbol{k},\boldsymbol{p})=\frac{1}{\rho^{3/2}}\left\langle\frac{\partial^{3}V}{\partial\alpha(\boldsymbol{q})\partial\beta(\boldsymbol{k})\partial\gamma(\boldsymbol{
 p})}\right\rangle_{\rho_{\mathcal{V}}}\sqrt{G_{\beta\gamma}(\boldsymbol{k},\boldsymbol{p})},
\end{equation}
\begin{multline}
 \label{fourth-order}
 \overset{(4)}{D}{}_{\alpha\beta\gamma\epsilon}(\boldsymbol{q},\boldsymbol{q}',\boldsymbol{k},\boldsymbol{k}')=\frac{1}{\rho^{2}}\left\langle\frac{\partial^{4}V}{\partial\alpha(\boldsymbol{q})\partial\beta(\boldsymbol{
 q}')\partial\gamma(\boldsymbol{k})\partial\epsilon(\boldsymbol{k}')}\right\rangle_{\rho_{\mathcal{V}}} \\ 
 \times \sqrt{G_{\alpha\beta}(\boldsymbol{q},\boldsymbol{k})G_{\gamma\epsilon}(\boldsymbol{q}',\boldsymbol{k}')}.
\end{multline}
The matrix $G_{\alpha\beta}(\boldsymbol{q},\boldsymbol{k})$ is defined as
\begin{equation}
 G_{\alpha\beta}(\boldsymbol{q},\boldsymbol{k})=\frac{F(0,\omega_{SCHA}^{\alpha}(\boldsymbol{q}),\omega_{SCHA}^{\beta}(\boldsymbol{k}))}{\omega_{SCHA}^{\alpha}(\boldsymbol{q})\omega_{SCHA}^{\beta}(\boldsymbol{k})},
\end{equation}
$F(0,\omega_{SCHA}^{\alpha}(\boldsymbol{q}),\omega_{SCHA}^{\beta}(\boldsymbol{k}))$ being the function defined in 
Eq. \eqref{raffaello-function}. We are interested in the corrections to the out-of-plane modes, therefore, we are
interested in the terms of the type
\begin{multline}
  D_{hh}^{corr}(-\boldsymbol{q},\boldsymbol{q})=\sum_{\gamma\delta\epsilon\zeta}\sum_{\boldsymbol{p}\boldsymbol{k}}\overset{(3)}{D}{}_{h\gamma\delta}(-\boldsymbol{q},\boldsymbol{p},
 \boldsymbol{q}-\boldsymbol{p})\times \\ \times  
 [1-\overset{(4)}{D}{}_{\gamma\delta\epsilon\zeta}(-\boldsymbol{p},\boldsymbol{p}-\boldsymbol{q},\boldsymbol{k},\boldsymbol{q}-\boldsymbol{k})]^{-1}\overset{(3)}{D}{}_{
 \epsilon\zeta h}(-\boldsymbol{k},\boldsymbol{k}-\boldsymbol{q},\boldsymbol{q}).
\end{multline}
By looking at Eq. \eqref{eq:ptential_da} we can see that only the terms of the type $\int_{\Omega}{d^{2}xC^{ijkl}\partial_{i}u_{j}\partial_{k}h\partial_{l}h}$ will contribute to the statistical average in
Eq. \eqref{third-order}. Therefore, Eq. \eqref{correction} can be rewritten as
\begin{multline}
\label{correction-simple}
 D_{hh}^{corr}(-\boldsymbol{q},\boldsymbol{q})=4\sum_{\alpha\beta}\sum_{\boldsymbol{p}\boldsymbol{k}}\overset{(3)}{D}{}_{hh\alpha}(-\boldsymbol{q},\boldsymbol{p},\boldsymbol{q}-\boldsymbol{p}) \times \\ \times  
 [ 1-\overset{(4)}{D}{}_{h\alpha h\beta}(-\boldsymbol{p},\boldsymbol{p}-\boldsymbol{q},\boldsymbol{k},\boldsymbol{q}-\boldsymbol{k})]^{-1}\overset{(3)}{D}{}_{h\beta h}(-\boldsymbol{k},\boldsymbol{k}-\boldsymbol{q},\boldsymbol{q}),
\end{multline}
where now the subindexes only run in $\alpha,\beta=u_{LA},u_{TA}$. Now, we can calculate the statistical averages
\begin{multline}
 \left\langle\frac{\partial^{3}V}{\partial h(\boldsymbol{k}_{1})\partial h(\boldsymbol{k}_{2})\partial u_{LA}(\boldsymbol{k}_{3})}\right\rangle_{\rho_{\mathcal{V}}}= \frac{1+\delta a}{\sqrt{\Omega}}\delta_{\boldsymbol{k}_{1}+\boldsymbol{
 k}_{2}+\boldsymbol{k}_{3},0}\times \\ \times  \left[\lambda|\boldsymbol{k}_{3}|\boldsymbol{k}_{1}\cdot\boldsymbol{k}_{2}+2\mu\frac{(\boldsymbol{k}_{3}\cdot\boldsymbol{k}_{1})(\boldsymbol{k}_{3}\cdot\boldsymbol{k}_{2})}{|\boldsymbol{k}_{3}|}\right],
\end{multline}
\begin{multline}
 \left\langle\frac{\partial^{3}V}{\partial h(\boldsymbol{k}_{1})\partial h(\boldsymbol{k}_{2})\partial u_{TA}(\boldsymbol{k}_{3})}\right\rangle_{\rho_{\mathcal{V}}}=\frac{\mu(1+\delta a)}{\sqrt{\Omega}}\delta_{\boldsymbol{k}_{
 1}+\boldsymbol{k}_{2}+\boldsymbol{k}_{3},0} \times \\ \times  \left[\frac{(\boldsymbol{k}_{3}\cdot\boldsymbol{k}_{1})(\boldsymbol{k}_{3\perp}\cdot\boldsymbol{k}_{2})+(\boldsymbol{k}_{3}\cdot\boldsymbol{k}_{2})(\boldsymbol{k}_{3\perp}\cdot
 \boldsymbol{k}_{1})}{|\boldsymbol{k}_{3}|}\right],
\end{multline}
\begin{multline}
\left\langle\frac{\partial^{4}V}{\partial h(\boldsymbol{k}_{1})\partial h(\boldsymbol{k}_{2})\partial u_{LA}(\boldsymbol{k}_{3})\partial u_{LA}(\boldsymbol{k}_{4})}\right\rangle_{\rho_{\mathcal{V}}}= \\ 
\frac{1}{\Omega}\delta_{\boldsymbol{
 k}_{1}+\boldsymbol{k}_{2}+\boldsymbol{k}_{3}+\boldsymbol{k}_{4},0}\frac{\boldsymbol{k}_{3}\cdot\boldsymbol{k}_{4}}{|\boldsymbol{k}_{3}||\boldsymbol{k}_{4}|}[\lambda(\boldsymbol{k}_{3}\cdot\boldsymbol{k}_{4})(\boldsymbol{k}_{
 1}\cdot\boldsymbol{k}_{2})+ \\ +\mu(\boldsymbol{k}_{3}\cdot\boldsymbol{k}_{1})(\boldsymbol{k}_{4}\cdot\boldsymbol{k}_{2})+\mu(\boldsymbol{k}_{3}\cdot\boldsymbol{k}_{2})(\boldsymbol{k}_{4}\cdot\boldsymbol{k}_{1})],
\end{multline}
\begin{multline}
 \left\langle\frac{\partial^{4}V}{\partial h(\boldsymbol{k}_{1})\partial h(\boldsymbol{k}_{2})\partial u_{TA}(\boldsymbol{k}_{3})\partial u_{TA}(\boldsymbol{k}_{4})}\right\rangle_{\rho_{\mathcal{V}}}= \\
 \frac{1}{\Omega}\delta_{\boldsymbol{
 k}_{1}+\boldsymbol{k}_{2}+\boldsymbol{k}_{3}+\boldsymbol{k}_{4},0}\frac{\boldsymbol{k}_{3\perp}\cdot\boldsymbol{k}_{4\perp}}{|\boldsymbol{k}_{3}||\boldsymbol{k}_{4}|}[\lambda(\boldsymbol{k}_{3}\cdot\boldsymbol{k}_{4})(\boldsymbol{k}_{
 1}\cdot\boldsymbol{k}_{2})+ \\ +\mu(\boldsymbol{k}_{3}\cdot\boldsymbol{k}_{1})(\boldsymbol{k}_{4}\cdot\boldsymbol{k}_{2})+\mu(\boldsymbol{k}_{3}\cdot\boldsymbol{k}_{2})(\boldsymbol{k}_{4}\cdot\boldsymbol{k}_{1})],
\end{multline}
and
\begin{multline}
 \left\langle\frac{\partial^{4}V}{\partial h(\boldsymbol{k}_{1})\partial h(\boldsymbol{k}_{2})\partial u_{LA}(\boldsymbol{k}_{3})\partial u_{TA}(\boldsymbol{k}_{4})}\right\rangle_{\rho_{\mathcal{V}}}= \\ 
 \frac{1}{\Omega}\delta_{\boldsymbol{
 k}_{1}+\boldsymbol{k}_{2}+\boldsymbol{k}_{3}+\boldsymbol{k}_{4},0}\frac{\boldsymbol{k}_{3}\cdot\boldsymbol{k}_{4\perp}}{|\boldsymbol{k}_{3}||\boldsymbol{k}_{4}|}[\lambda(\boldsymbol{k}_{3}\cdot\boldsymbol{k}_{4})(\boldsymbol{k}_{
 1}\cdot\boldsymbol{k}_{2})+ \\ +\mu(\boldsymbol{k}_{3}\cdot\boldsymbol{k}_{1})(\boldsymbol{k}_{4}\cdot\boldsymbol{k}_{2})+\mu(\boldsymbol{k}_{3}\cdot\boldsymbol{k}_{2})(\boldsymbol{k}_{4}\cdot\boldsymbol{k}_{1})].
\end{multline}
The equations cannot be further simplified but we have all the ingredients to calculate them numerically. We have checked numerically that, as in the atomistic case, the contribution of $\overset{(4)}{\boldsymbol{D}}$ is completely negligible. To show that we Taylor expand Eq. \eqref{scha-se}
\begin{multline}
 \boldsymbol{\Pi}(z)=\mathbf{M}^{-\frac{1}{2}}\overset{(3)}{\mathbf{\Phi}}\mathbf{\Lambda}(z)[\mathbf{1}-\overset{(4)}{\mathbf{\Phi}}\mathbf{\Lambda}(z)]^{-1}
 \overset{(3)}{\mathbf{\Phi}}\mathbf{M}^{-\frac{1}{2}} \simeq \\ \simeq
 \mathbf{M}^{-\frac{1}{2}}\overset{(3)}{\mathbf{\Phi}}\mathbf{\Lambda}(z)\overset{(3)}{\mathbf{\Phi}}\mathbf{M}^{-\frac{1}{2}}+\mathbf{M}^{-\frac{1}{2}}\overset{(3)}{\mathbf{\Phi}}\mathbf{\Lambda}(z)\overset{(4)}{\mathbf{\Phi}}\mathbf{\Lambda}(z)\overset{(3)}{\mathbf{\Phi}}\mathbf{M}^{-\frac{1}{2}},
\end{multline}
and we calculate the contribution of the term containing the fourth-order tensor to the linewidth. We also calculate the spectral function with and without including the frequency dependence of the self energy. We show the results in Extended Data Fig. \ref{lw_membrane}. The figure clearly shows that the contribution of the fourth-order tensor is at least one order of magnitude smaller than the main term, justifying the bubble approximation of the self-energy,  and, what it is more important, it also decays as momentum decreases. The figure also shows that the Lorentzian approximation is justified for the acoustic modes in graphene. \\

By neglecting the fourth-order terms containing in-plane displacement fields in Eq. \ref{full-potential}, the SCHA can be applied analytically in this model. The SCHA equations simplify to
\begin{equation}
\label{scha-tommaso1}
\delta a=-\frac{1}{4\Omega}\sum_{\boldsymbol{q}}|\boldsymbol{q}|^{2}g[\omega_{SCHA}^{(h)}(\boldsymbol{q})],
\end{equation}
\begin{multline}
\label{scha-tommaso2}
\Phi_{SCHA}^{(h)}(\boldsymbol{q})=\kappa|\boldsymbol{q}|^{4}+2\delta a(\lambda+\mu)|\boldsymbol{q}|^{2}+\\ +\frac{\lambda+2\mu}{2\Omega}\sum_{\boldsymbol{k}}g[\omega_{SCHA}^{(LA)}(\boldsymbol{k})][|\boldsymbol{q}|^{2}|\boldsymbol{k}|^{2}+2(\boldsymbol{q}\cdot\boldsymbol{k})^{2}].    
\end{multline}
By inserting Eq. \eqref{scha-tommaso1} in Eq. \eqref{scha-tommaso2} and considering the infinite volume limit ($\Omega\rightarrow\infty$), we obtain
\begin{equation}
\label{tommaso1}
\Phi_{SCHA}^{(h)}(\boldsymbol{q})=\kappa|\boldsymbol{q}|^{4}+\gamma|\boldsymbol{q}|^{2},    
\end{equation}
where $\gamma$ is given by the solution of
\begin{equation}
\label{tommaso2}
\gamma=\gamma\frac{\lambda+3\mu}{16\pi\kappa\sqrt{\rho\kappa}}\int_{0}^{\Lambda\sqrt{\kappa/\gamma}}ds\frac{s^{2}coth[\gamma s\sqrt{1+s^{2}}/(2T\sqrt{\rho\kappa})]}{\sqrt{1+s^{2}}}.
\end{equation}
$\Lambda$ is an ultraviolet cutoff that avoids divergencies.
Eqs. \ref{tommaso1} and \ref{tommaso2} show that the dispersion of the SCHA auxiliary ZA modes is linear. By calculating the correction for getting the physical phonons in the static approach in Eq. \ref{correction-simple} (in this case the fourth-order tensor is $0$) the result is
\begin{equation}
\Phi_{F}^{(h)}(\boldsymbol{q})=\kappa|\boldsymbol{q}|^{4}+(\gamma-\sigma)|\boldsymbol{q}|^{2}+O(|\boldsymbol{q}|^{4}),    
\end{equation}
where at $T=0$ K
\begin{equation}
\sigma=\frac{\rho\sqrt{\gamma}}{8\pi\kappa^{3/2}}\sum_{\alpha=LA,TA}v_{\alpha}f(\Lambda\sqrt{\kappa/\gamma},v_{\alpha}\sqrt{\rho/\gamma}),    
\end{equation}
with
\begin{equation}
f(x,y)=\int_{0}^{x}ds\frac{s^{2}}{\sqrt{1+s^{2}}[\sqrt{1+s^{2}}+y]}.
\end{equation}
By setting the ultraviolet cutoff to the value of the Debye momentum, $\Lambda=\sqrt{\frac{8\pi}{3^{1/2}a_{0}}}=1.55\AA$, we obtain $1-\sigma/\gamma=20\%$. This means that the linear component of the physical frequencies turns out to be a factor of $40\%$ smaller than the one of the SCHA auxiliary frequency. The non zero linear term in the physical frequencies appears because neglecting the fourth-order terms including in-plane displacements breaks the rotational invariance of the potential.\\

\textbf{The equal time height-height correlation function within SCHA}.
Within an interacting picture, the ensemble average of any displacement-displacement correlation function is given by the following equal time Green function (we use $\hbar=k_{B}=1$):
\begin{equation}
	\sqrt{M_aM_b}\langle u_a u_b \rangle = G^{ab}(\tau=0^{+})=-T\sum_{n}G^{ab}(i \Omega_n),
\end{equation}
where $G^{ab}(i \Omega_n)$ is the  SCHA Green function in frequency domain for the variable $\sqrt{M_a}$($R^a-\mathcal{R}^{a}_{eq}$) defined in Eq. \eqref{green-dyson} and $\Omega_n= 2\pi T n$ are the bosonic Matsubara's frequencies. $\mathcal{R}^{a}_{eq}$ are the centroid positions that minimize the SCHA free energy.\\

The summation has to be done via the Lehmann representation:
\begin{equation}
	\sqrt{M_aM_b}\langle u_a u_b \rangle =-T\sum_{n}G^{ab}(i \Omega_n)= \int_{-\infty}^{\infty} \frac{d \omega}{2\pi} \; \sigma (\omega) n_{B}(\omega)
\end{equation}
being $\sigma(\omega)$  the spectral function of the Green function: $\sigma(\omega)=-2 \text{Im}[G(\omega+i0^{+})]$. Retaining only the first term of the dynamical SCHA self energy (bubble aproximation) and neglecting the mode-mixing, the spectral function resembles a superposition of Lorentzians, but with frequency dependent shifts and widths. When the quasiparticle picture is valid after the inclusion of anharmonicity, the spectral function can actually be expressed as a superposition of Lorentzians:  
\begin{multline}\label{eqn: sf}
\sigma (\omega)= \sum_{\mu} \epsilon_{\mu}^a \epsilon_{\mu}^b \Bigg( \frac{1}{\omega} \Bigg[ \frac{\Gamma_{\mu}}{(\omega-\Theta_{\mu})^2+ (\Gamma_{\mu}) ^2}+\\+\frac{\Gamma_{\mu}}{(\omega+\Theta_{\mu})^2+ (\Gamma_{\mu}) ^2} \Bigg]\Bigg),
\end{multline}
where $\Theta_{\mu}$ is the frequency of the SCHA quasiparticle in the Lorentzian approximation and $\Gamma_{\mu}$ its HWHM linewidth.\\

We can avoid divergences in the integral by redefining the sum as
\begin{equation} \label{eqn: Tom}
	\sqrt{M_aM_b} \langle u_a u_b \rangle =-TG^{ab}(0)+ \int_{-\infty}^{\infty} \frac{d \omega}{2\pi} \;  \sigma (\omega) \left[n_{B}(\omega)-\frac{T}{\omega}\right].
\end{equation}
Regarding the first term in the sum, the static limit of the Green function corresponds to the inverse of the free energy dynamical matrix:
\begin{equation}\label{eqn: static}
	G^{ab}(i\Omega_n=0)=-[D^{(F)}]^{-1}_{ab}= \sum_{\mu} \epsilon_{\mu}^a \epsilon_{\mu}^b \left(-\frac{1}{\Omega_{\mu}^2}\right),
\end{equation}
where $\Omega_{\mu}$ are again the frequencies of the free energy phonons.\\

Inserting Eqs. \eqref{eqn: static} and \eqref{eqn: sf} in Eq. \eqref{eqn: Tom}:
\begin{multline} 
	\sqrt{M_aM_b} \langle u_a u_b \rangle =\sum_{\mu} \epsilon_{\mu}^a \epsilon_{\mu}^b \Bigg( \frac{T}{\Omega_{\mu}^2}+ \int_{-\infty}^{\infty} \frac{d \omega}{2\pi} \times \\ \times \Big( \frac{1}{\omega}\left[\frac{\Gamma_{\mu}}{(\omega-\Theta_{\mu})^2+ (\Gamma_{\mu}) ^2}+\frac{\Gamma_{\mu}}{(\omega+\Theta_{\mu})^2+ (\Gamma_{\mu}) ^2}\right] \Big)   \times \\ \times \left[n_{B}(\omega)-\frac{T}{\omega}\right] \Bigg) .
\end{multline}
This integral can be simplified when the phonon-phonon linewidth tends to zero. For those cases, the Lorentzian representation of the Dirac delta function can be used:
\begin{equation}
	\delta(x)=\frac{1}{\pi} \lim_{\epsilon \to 0^{+}} \frac{\epsilon}{x^2+\epsilon^2}.
\end{equation}
Then,
\begin{multline} 
	\sqrt{M_aM_b}\langle u_a u_b \rangle =\sum_{\mu} \epsilon_{\mu}^a \epsilon_{\mu}^b \Bigg( \frac{T}{\Omega_{\mu}^2}+ \frac{1}{2}\int_{-\infty}^{\infty} d\omega  \times \\ \times \left(\frac{1}{\omega} \times \left[\delta(\omega-\Theta_{\mu})+\delta(\omega+\Theta_{\mu}) \right] \right)   \left[n_{B}(\omega)-\frac{T}{\omega}\right] \Bigg).
\end{multline}
And
\begin{equation*} 
	\langle u_a u_b \rangle =\frac{\sum_{\mu} \epsilon_{\mu}^a \epsilon_{\mu}^b}{\sqrt{M_aM_b}} \left(\frac{T}{\Omega_{\mu}^2}+ \frac{-\hbar n_{B}(-\Theta_{\mu})+\hbar n_{B}(\Theta_{\mu})}{2\Theta_{\mu}}-\frac{T}{\Theta_{\mu}^2}\right).
\end{equation*}
Finally, when free energy Hessian phonons (physical phonons in the static approach) and physical ones are nearly identical ($\Omega_{\mu}^2 \approx \Theta_{\mu}^2$), we recover the formula of the non-interacting case but evaluated with free energy Hessian (equivalently, physical) phonons:
\begin{equation}
\begin{split}  
	\langle u_a u_b \rangle &= \frac{\sum_{\mu} \epsilon_{\mu}^a \epsilon_{\mu}^b}{\sqrt{M_aM_b}} \left[ \frac{ \left(n_B[\Omega_\mu]-n_B[-\Omega_\mu]\right)}{2\Omega_\mu}\right]= \\  &= \frac{\sum_{\mu} \epsilon_{\mu}^a \epsilon_{\mu}^b}{\sqrt{M_aM_b}} \left[ \frac{\left(1+2n_B[\Omega_\mu]\right)}{2\Omega_\mu}\right].
\end{split}
\end{equation}
In the case of the membrane model, the displacement-displacement correlation function is:
\begin{equation}
	\langle u_a(\textbf{x})u_b(\textbf{x}') \rangle = \frac{\sum_{\mu} \epsilon_{\mu}^a (\textbf{x}) \epsilon_{\mu}^b (\textbf{x}')}{\rho} \left[ \frac{\left(1+2n_B[\Omega_\mu]\right)}{2\Omega_\mu}\right],
\end{equation}
where $a$ and $b$ are the Cartesian indexes and $\mu = h,LA,TA$ in this case. Essentially, discrete magnitudes are now continuous, while the individual atomic masses $M_a$ and $M_b$ are replaced by the mass density of the membrane $\rho$. The corresponding Fourier transform is given by
\begin{equation}
	\langle u_a(\textbf{q})u_b(\textbf{k}) \rangle = \delta_{\textbf{q},-\textbf{k}}\frac{\sum_{\mu} \epsilon_{\mu}^a (\textbf{q}) \epsilon_{\mu}^b (-\textbf{q})}{\rho} \left[ \frac{\left(1+2n_B[\Omega_\mu(\textbf{q})]\right)}{2\Omega_\mu(\textbf{q})}\right].
\end{equation}
We are particularly interested on the Fourier transform of the out-of-plane correlation function. As in the membrane model $ZA$ is the only mode with an out-of-plane component, we finally obtain:
\begin{equation} 
		\langle |h(\textbf{q})|^2 \rangle = \frac{\left(1+2n_B[\Omega_{ZA}(\textbf{q})]\right)}{2\rho\Omega_{ZA}(\textbf{q})},
\end{equation}
which is the formula implemented along this article to obtain the Fourier transform of the height-height correlation function.\\

\begin{figure*}[ht]
\centering
\includegraphics[width=0.8\linewidth]{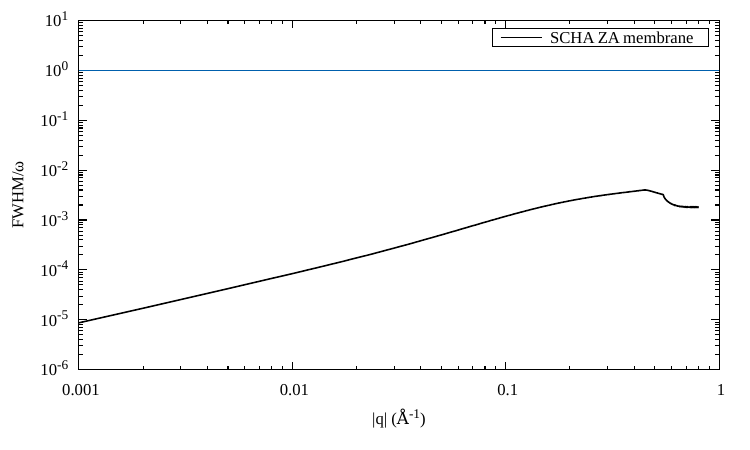}
\caption{Linewidth (full width half maximum) of ZA phonon mode divided by its frequency at 300K calculated within the membrane model. }
\label{checkv4-membrane.}
\end{figure*}

Nearly all the approximations taken in this mathematical derivation have been proved for the graphene throughout this article. The only task left is showing that the linewidth of the ZA mode is as small as the ones corresponding to the in-plane phonon modes, which is indeed true as shown in Extended Data Fig. \ref{checkv4-membrane.}.

\textbf{Extra calculations of the equal time  height-height correlation function}.
The out-of-plane correlation function is governed by the bosonic occupation factor. Quantum correlations appear for those flexural modes that are barely occupied thermally, that is, in those modes which quantum zero-point energy is bigger than the thermal energy:
\begin{equation}
		 \frac{1}{2}  \hbar \omega_{ZA}(\textbf{q}) > K_BT < => q_T > \left(\sqrt{\frac{\rho}{\kappa}}\frac{2K_BT}{\hbar}\right).
\end{equation}
Decreasing the temperature and/or increasing the wavelength favours the emergence of quantum correlations ~\cite{pimc}.\\

\begin{figure*}[h!]
\includegraphics[width=0.7\linewidth]{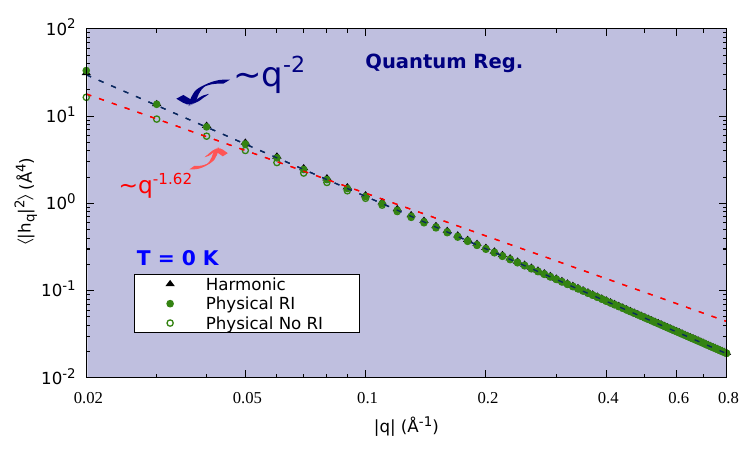}
\centering
\caption{Fourier transform of the height-height correlation function at 0 K in the membrane model evaluated at different levels of approximation: harmonic (black dots), anharmonic rotationally invariant (RI) result (green filled dots) and anharmonic no rotationally invariant (No RI)  result (green empty dots). The dashed lines correspond to the linear fitting in each case.}
\label{hh0}
\end{figure*}

In this subsection we provide extra calculations analyzing the extreme cases at 0 K and 300 K.
At null temperature there is no phonon mode thermally occupied, but all of them fluctuate due to quantum zero-point motion. The height-height correlation function shows then a fully quantum behaviour, with no crossover to a classical regime as shown in Extended Data Fig. \ref{hh0}. The harmonic and anharmonic rotational invariant results yield the same exponents  due  to  their  quadratic dispersion: $\langle |h(\textbf{q})|^2 \rangle  \sim q^{-2}$. The anharmonic non rotational invariant phonons are quadratic in the short wavelength limit, but they are linearized in the long wavelength limit with $\langle |h(\textbf{q})|^2 \rangle \sim \Omega_{ZA}(\textbf{q})^{-1}\sim q^{-1.62} $. This exponent coincides with the one obtained in the self consistent screening approximation (SCSA), which scale as $q^{\nu}$ with $\nu \sim 1.6$~\cite{roldan2011suppression}.

At 300 K the classic to quantum crossover occurs at 1.18 $\AA^{-1}$, so that all the modes are largely occupied in the $q$ range in which we have focused our analysis. Thermal fluctuations rule and the height-height correlation function shows a classical behaviour. Again, the quadratic dispersion of the harmonic and anharmonic rotaional invariant results is behind the exponent of the correlation function, which is now of $\langle |h(\textbf{q})|^2 \rangle  \sim q^{-4}$ as predicted by classical statistics. The linearization of the anharmonic phonons in the long wavelength limit when the rotational symmetry is broken makes us recover the exponent obtained in classical references in the literature.

\begin{figure*}[h]
\centering
\includegraphics[scale=0.65]{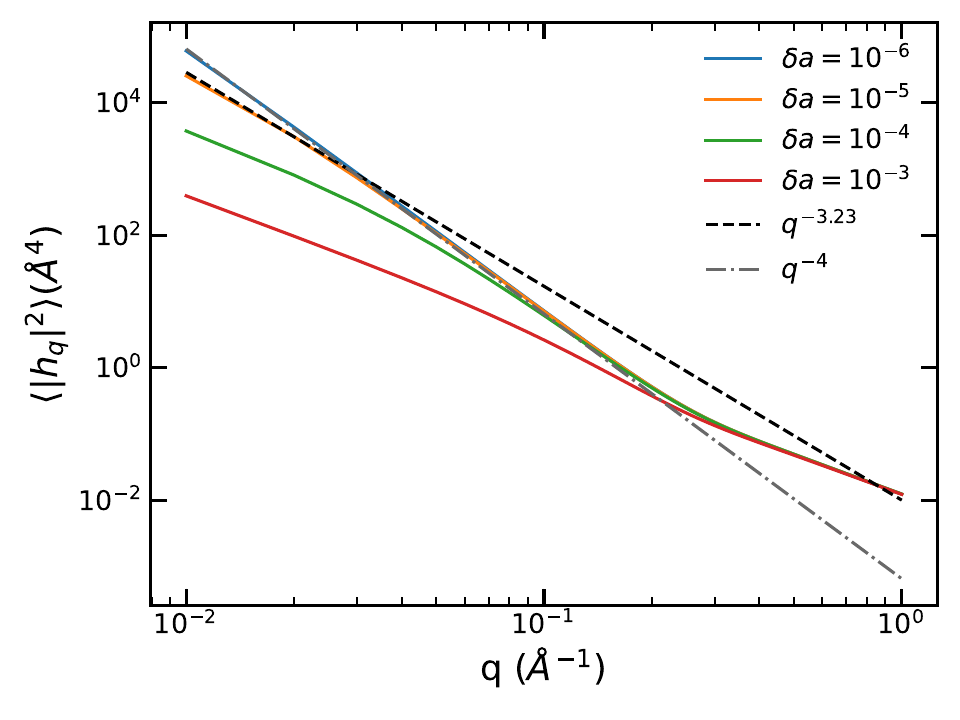}
\caption{This figure represents the value of $\braket{|h(\mathbf{q})|^2}$ as a function of the biaxial strain $\delta a$. Impressively, the behavior for small $q$ deviates from the $q^{-4}$ law even for very small strains, e.g. $\delta a = 10^{-5}$. }
\label{h2}
\end{figure*}

\textbf{Dependence of the ZA frequency on the strain}. 
To assess the significance of small strains on the behavior of the height-height correlation function, we formulate a simple harmonic model that describes the relationship between the ZA frequency and the biaxial strain $\delta a$. In Eq. \eqref{eq:ptential_da} the only second-order term involving $h$ is $\delta a(\lambda+\mu)\int_{\Omega}d^2x\ \partial_kh \partial_kh$.
Consequently, the modified harmonic potential energy for $h$ due to strain can be expressed as
\begin{equation}
    U_{\delta a} = \frac{1}{2}\Bigl[\int_{\Omega} d^2x\ \kappa (\partial^2h)^2 + 2\delta a(\lambda+\mu)\int_{\Omega}d^2x\ \partial_kh \partial_kh\Bigr]
\end{equation}
whose diagonalization leads to 
\begin{equation}\label{omega_eps}
    \omega_{ZA}(q)= \sqrt{\frac{2(\lambda+\mu)\delta aq^2+\kappa q^4}{\rho}}\ . 
\end{equation}
We plug Eq. \eqref{omega_eps} in the equation for the heigh-height correlation function in the main text (Eq. \eqref{eq:hhq})
and calculate explicitly $\langle |h(\mathbf{q})|^2\rangle$ at $T=12.5$ K. The result is shown in Extended Data Fig. \ref{h2}. A strain as small as $\delta a = 10^{-5}$ can deviate the ripples amplitude from the $q^{-4}$ law lowering it to $q^{-3.23}$.

\section*{Data availability}

All the data generated in this work is available upon request from I.E.

\section*{Code availability}

The atomistic calculations of the SCHA theory are performed with the SSCHA code. This code is open source and can be downloaded from \url{www.sscha.eu}. The calculations of the SCHA in the membrane model are performed with an in-house code.

\section*{Acknowledgements}

We would like to thank Francisco Guinea for useful conversations. Financial support was provided by the Spanish Ministry of Economy and Competitiveness (FIS2016-76617-P);  the Spanish Ministry of Science and Innovation (Grant No. PID2019-105488GB-I00); the Department of Education, Universities and Research of the Basque Government and the University of the Basque Country (IT1707-22 and IT1527-22);
and the European Commission under the Graphene Flagship, Core 3, grant number 881603.
U.A. is also thankful to the Material Physics Center for a predoctoral fellowship. 
J.D. thanks the Department of Education of the Basque Government for a predoctoral fellowship (Grant No. PRE-2020-1-0220).
Computer facilities were provided by the Donostia International Physics Center (DIPC).

\section*{Author contributions}

U.A. performed the atomistic calculations, while U.A., J.D., and T.C. performed the calculations on the membrane and developed the theoretical adaptation of the SCHA theory to the membrane. I.E. and F.M. supervised the full project. The manuscript was written by U.A., J.D., and I.E. with input from all authors. 

\section*{Competing interests}

The authors declare no competing financial interests.
Correspondence and requests for materials should be addressed
to I.E. (ion.errea@ehu.eus).

%---------------------------------------------------------
% 
% EXTENDED DATA FIGURES
%

\newpage

\newpage

\newpage

\newpage

\newpage

\newpage

\newpage    

\newpage

\newpage

\begin{figure*}[h!]
\includegraphics[width=0.7\linewidth]{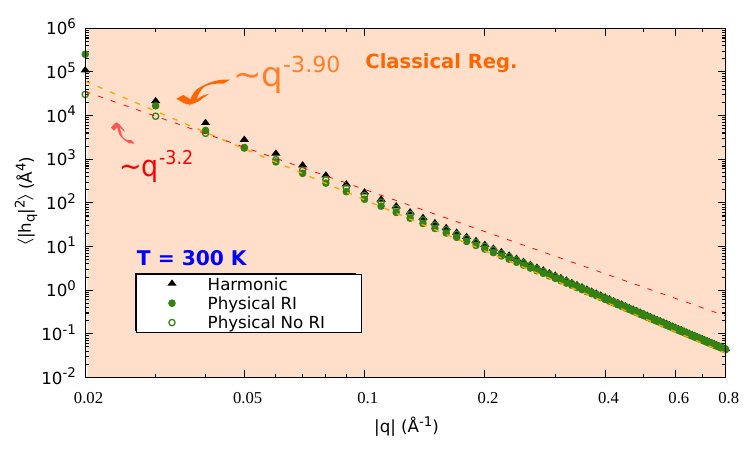}
\centering
\caption{Fourier transform of the height-height correlation function at 300 K in the membrane model evaluated at different levels of approximation: harmonic (black dots), anharmonic RI result (green filled dots) and anharmonic No RI result (green empty dots). The dashed lines correspond to the linear fitting in each case.}
\label{hh300}
\end{figure*}

\newpage

\end{document}